\documentclass[runningheads, a4paper]{llncs}
\usepackage{amssymb}
\usepackage{url}

\usepackage{geometry}
\usepackage[english]{babel} 
\usepackage[utf8]{inputenc}
\usepackage[T1]{fontenc}

\usepackage{xspace}

\usepackage{xcolor}
\usepackage{wrapfig}
\usepackage{multirow}
\usepackage{pdflscape}  

\usepackage{mathtools} 
\usepackage{booktabs} 
\usepackage{stmaryrd}
\usepackage{graphics,color}
\usepackage{graphicx}
\graphicspath{{./figures/}}

\usepackage{enumerate}
\usepackage{listings}

\lstdefinelanguage{prog}{morekeywords={let,lock,unlock,fn,fun,in,stop,new,spawn}}

\usepackage{array}
\newcolumntype{L}[1]{>{\raggedright\let\newline\\\arraybackslash\hspace{0pt}}m{#1}}
\newcolumntype{C}[1]{>{\centering\let\newline\\\arraybackslash\hspace{0pt}}m{#1}}
\newcolumntype{R}[1]{>{\raggedleft\let\newline\\\arraybackslash\hspace{0pt}}m{#1}}

\def\codesize{\fontsize{10}{11}}

\makeatletter
\newenvironment{btHighlight}[1][]
{\begingroup\tikzset{bt@Highlight@par/.style={#1}}\begin{lrbox}{\@tempboxa}}
{\end{lrbox}\bt@HL@box[bt@Highlight@par]{\@tempboxa}\endgroup}

\newcommand\btHL[1][]{%
  \begin{btHighlight}[#1]\bgroup\aftergroup\bt@HL@endenv%
}
\def\bt@HL@endenv{%
  \end{btHighlight}%
  \egroup
}
\newcommand{\bt@HL@box}[2][]{%
  \tikz[#1]{%
    \pgfpathrectangle{\pgfpoint{1pt}{0pt}}{\pgfpoint{\wd #2}{\ht #2}}%
    \pgfusepath{use as bounding box}%
    \node[anchor=base west, fill=gray!20,outer sep=0pt,inner xsep=1pt, inner ysep=0pt, rounded corners=3pt, minimum height=\ht\strutbox+1pt,#1]{\raisebox{1pt}{\strut}\strut\usebox{#2}};
  }%
}
\makeatother

\newcommand{\red}[1] {\textcolor{red}{#1}}
\newcommand{\blue}[1] {\textcolor{blue}{#1}}

\newcommand{\rn}[1]{\mbox{\textsc{#1}}}
\newcommand{\many}[1]{\overline{#1}} 
\newcommand{\hsep} {,\;}                         
\newcommand{\lsep}{\bullet}
\newcommand{\config}                               {\mathit{Config}}
\newcommand{\memory}                    {M}

\newcommand{\cid}                             {cid}
\newcommand{\core}                          {CR}
\newcommand{\levid}                             {caid}

\newcommand{\cache}                       {\mathit{Ca}}
\newcommand{\flag}				{\mathit{last}}
\newcommand{\status}                       {\mathit{st}}
\newcommand{\mo}                               {\mathit{mo}}
\newcommand{\sh}                               {\mathit{sh}}
\newcommand{\inv}                               {\mathit{inv}}
\newcommand{\task}                          {\mathit{rst}}
\newcommand{\stask}                        {\mathit{dap}}
\newcommand{\datainst}                        {\mathit{dst}}

\newcommand{\key}[1]{\mbox{\sffamily\bfseries\color{kwcolor} #1}}
\newcommand{\none}{\varepsilon}
\newcommand{\blocks}[1]				{\key{fetchBl}(#1)}
\newcommand{\fetchwait}[2]	{\key{fetchW}(#1,#2)}

\newcommand{\flushs}[1]                          {\key{flush}(#1)}
\newcommand{\fetchs}[1]                          {\key{fetch}(#1)}
\newcommand{\reads}[1]                            {\key{read}(#1)}
\newcommand{\readBs}[1]                            {\key{readBl}(#1)}
\newcommand{\writes}[1]               {\key{write}(#1)}
\newcommand{\writeBs}[1]               {\key{writeBl}(#1)}

\newcommand{\pr}[2]                               {#2}

\newcommand{\sendR}[1]{!\mathit{Rd(#1)}}
\newcommand{\recR}[1]{?\mathit{Rd(#1)}}
\newcommand{\sendRX}[1]{!\mathit{RdX(#1)}}
\newcommand{\recRX}[1]{?\mathit{RdX(#1)}}

\newcommand{\statusfunc}[2]                               {\mathit{status}(#1,#2)}
\newcommand{\selectfunc}[2]                               {\mathit{select}(#1,#2)}

\newcommand{\first}[1]				{\mathit{first}(#1)}
\newcommand{\last}[1]				{\mathit{last}(#1)}
\newcommand{\coreid}[1]				{\mathit{cid}(#1)}
\newcommand{\levno}[1]				{\mathit{lid}(#1)}

\newcommand{\dom}[1]                             {\mathit{dom}(#1)}

\newcommand{\toL}[1] {\xrightarrow{#1}}

\newcommand{\setto}[2]                    {[#1\mathop{\mapsto}#2]} 

\newcommand{\msunion}{+}

\newcommand{\crule}[1]{\blue{#1}}

\newcommand{\hist}[1]{}

\newcommand{\oid}{\mathit{oid}}

\definecolor{airforceblue}{rgb}{0.36, 0.54, 0.66}
\definecolor{bittersweet}{rgb}{1.0, 0.44, 0.37}
\definecolor{brilliantlavender}{rgb}{0.96, 0.73, 1.0}
\newcommand{\hlc}[1]{%
  \colorbox{airforceblue!25}{$\displaystyle#1$}}

\newcommand{\hlcc}[1]{%
  \colorbox{bittersweet!25}{$\displaystyle#1$}}

\newcommand{\ncondrule}[3]{ 
  \begin{array}{c} 
    \textsc{ (\blue{#1})} \\[1pt] 
    #2 \\[1pt] 
    \hline\\[-7pt]
    #3 
  \end{array} }

\newcounter{magicrownumbers}

\lstdefinelanguage{ABS}{keywords=
{assert, switch, when, newclass,newinterface,update,foreach,null,this,thisDC,dyndelta,new,data,type,def,case,case2,of,cog,class,interface,extends,implements,if,else,await,get,local,Fut,return,skip,while,module,duration,duration2,now,deadline,import, export, uses, from,
  suspend,delta,adds,modifies,removes,original,then,productline,features ,corefeatures,optionalfeatures,after,when,product,hasAttribute,hasMethod,root,extension,group,allof,oneof,require,exclude,original,ifin,ifout,opt,spawn,newgroup,acquire,in,except,joins,leaves,as,subtypeOf}, 
sensitive=true, comment=[l]{//}, morecomment=[s]{/*}{*/}, morestring=[b]"}

\definecolor{codegreen}{rgb}{0,0.6,0}
\definecolor{codegray}{rgb}{0.5,0.5,0.5}
\definecolor{kwcolor}{rgb}{0.58,0,0.82} 
\definecolor{backcolour}{rgb}{0.95,0.95,0.92}

\newcommand{\Abs}[1]{\lstinline[language=ABS,columns=fullflexible,mathescape=true,inputencoding=latin1,extendedchars,keywordstyle=\bf\sffamily\codesize,basicstyle=\sffamily\codesize]!#1!}

\newenvironment{absinline}[1][]{%
\lstset{language=ABS,
		 showstringspaces=false,
		 columns=fullflexible,mathescape=true,%
         keywordstyle=\bf\sffamily,commentstyle=\sl\sffamily,%
         basicstyle=\normalsize\sffamily,inputencoding=latin1, 
         extendedchars,xleftmargin=2em,#1}}
{}

\lstdefinestyle{absstyle}{
language=ABS,columns=fullflexible,
 		   mathescape=true,%
 		   showstringspaces=false,%
keywordstyle=\bf\sffamily,
numberstyle=\scriptsize,
commentstyle=\sl\sffamily,%
basicstyle=\footnotesize\sffamily,
inputencoding=latin1, 
extendedchars,xleftmargin=2pt
}

\lstnewenvironment{absexample}{
\noindent 
\lstset{style=absstyle,
basicstyle=\myttsize\sffamily,
keywordstyle=\bfseries\color{kwcolor},
framerule=1pt,
backgroundcolor=\color{backcolour},
rulecolor=\color{kwcolor!80!black},
frame=tblr,
xleftmargin=4pt,
xrightmargin=6pt,
}}
{}

\lstnewenvironment{absexamplen}[1]{
\lstset{style=absstyle,
basicstyle=\myttsize\sffamily,
keywordstyle=\bfseries\color{kwcolor},
framerule=1pt,
backgroundcolor=\color{backcolour},
rulecolor=\color{kwcolor!80!black},
frame=tblr,
xleftmargin=4pt,
xrightmargin=6pt,#1
}}
{}

\lstnewenvironment{absexamplesm}{
\noindent 
\lstset{style=absstyle, numbers=left,numberstyle=\tiny\texttt,xleftmargin=15pt,
basicstyle=\small\sffamily,
keywordstyle=\bfseries\color{kwcolor},
framerule=1pt,
backgroundcolor=\color{backcolour},
rulecolor=\color{kwcolor!80!black},
frame=tblr,
xrightmargin=6pt,
}}
{}

\let\origthelstnumber\thelstnumber
\makeatletter
\newcommand*\Suppressnumber{%
  \lst@AddToHook{OnNewLine}{%
    \let\thelstnumber\relax%
     \advance\c@lstnumber-\@ne\relax%
    }%
}

\newcommand*\Reactivatenumber[1]{%
  \setcounter{lstnumber}{\numexpr#1-1\relax}
  \lst@AddToHook{OnNewLine}{%
   \let\thelstnumber\origthelstnumber%
   \refstepcounter{lstnumber}%
  }%
}
\makeatother

\def\myttsize{\fontsize{9}{9.5}}
\lstdefinelanguage{MINICORE}{keywords=
{task, main , read , write , spawn }, sensitive=true, comment=[l]{//}, morecomment=[s]{/*}{*/}, morestring=[b]"}

\lstdefinestyle{minicorestyle}{
language=MINICORE,columns=fullflexible,numbers=none,
 		   mathescape=true,%
 		   showstringspaces=false,%
keywordstyle=\bf\sffamily,
commentstyle=\sl\sffamily,%
basicstyle=\footnotesize\sffamily,
inputencoding=latin1, 
extendedchars,xleftmargin=2em
}

\lstnewenvironment{minicoreexample}{
~\\
\noindent 
\small{\textcolor{Gray!80!black}{\bfseries\sffamily{Example:}}}\vspace{-4pt}
\lstset{style=minicorestyle,
basicstyle=\\sffamily,
framerule=1pt,
frame=tblr,
xleftmargin=4pt,
}}
{}

\lstnewenvironment{minicoreexamplen}[1][]{
\lstset{style=minicorestyle,
basicstyle=\myttsize\sffamily,
framerule=1pt,tabsize=2,
frame=tblr,
captionpos=b,
xleftmargin=4pt,#1
}}
{}

\newcommand{\onelettername}[1]{#1}
\title{Proving Correctness of Parallel  Implementations \\ of Transition System  Specifications} 
\author{Frank S. de Boer \inst{1} 
 \and Einar Broch  Johnsen \inst{2} \and  Violet Ka I
  Pun \inst{3} \and S. Lizeth Tapia Tarifa \inst{2} }
\institute{
CWI, The Netherlands \\ 
\email{F.S.de.Boer@cwi.nl}
\and
  Department of Informatics, University of Oslo, Norway\\
  \email{\{einarj,sltarifa\}@ifi.uio.no} 
\and
Department of Computing, Western Norway University of Applied Sciences,  Norway   \\
  \email{Violet.Ka.I.Pun@hvl.no}
  }
  
  \titlerunning{Proving Correctness of TSS}
  \authorrunning{F.~S. de Boer \emph{et al.}}

\begin{document}

 \maketitle 
 
\begin{abstract}
  The overall problem addressed in this paper is the long-standing
  problem of program correctness, and in particular programs
  that describe systems of parallel executing processes.  We propose a
  new method for proving correctness of parallel implementations of
  high-level transition system specifications. The implementation
  language underlying the method is based on the model of active (or
  concurrent) objects.  The method defines correctness in terms of a
  simulation relation between the transition system which specifies
  the program semantics and the transition system that is described by
  the correctness specification.  The simulation relation itself
  abstracts from the fine-grained interleaving of parallel processes
  by exploiting a global confluence property of the particular model
  of active objects considered in this paper.

  As a proof-of-concept we apply our method to the correctness of a
  parallel simulator of multicore memory systems.
\end{abstract}

\section{Introduction}
\label{sec:introduction}

A long-standing challenge in Computer Science is the formal
specification and verification of programs, notably that of parallel
programs, e.g., multi-threaded Java programs which
give rise to complex fine-grained interleaving of the parallel
executing processes and interaction (locking) mechanisms.

Roughly, we can distinguish between logic and semantic based methods
for establishing program correctness.  Methods that are based on logic
use assertions to express behavioral properties and generate proof
conditions for their validation, which are usually discharged by
interactive theorem proving.  These methods are applicable to
infinite-state systems and to actual programs used in practice (see
for example \cite{GouwBBHRS19} for the verification of a corrected
version of the TimSort sorting program of the Java Collections
Framework).  One of the main complexities of the use of logics is due
to the complexity of the specification of invariant properties and the
interactive use of a theorem prover.  On the other hand,
model-checking is based on an automated state-space exploration of the
program execution.  This in general is restricted to finite-state
systems and requires suitable abstraction techniques to master the
state-space explosion problem.
 
The main contribution of this paper is a new method which supports the
specification of an abstraction of the overall behavior of a parallel
program in terms of a \emph{transition system specification} (TSS, for
short) \cite{G19,bol96jacm,groote92ic}.  Verifying that a parallel program
satisfies such a correctness specification then involves establishing
a \emph{simulation relation} between the transition system describing
the semantics of the parallel program and the system described by the
TSS.

Our method supports  a general approach to proving the correctness of
parallel programs in two steps:
\begin{enumerate}
\item Verify global behavioral properties using a high-level formal model which
  abstracts from the complexity of the concurrency model of the target
  language to support inductive proofs of global properties.
\item Justify the correctness of the parallel implementation in the
  target language with respect to the high-level model in terms of a
  simulation relation.
\end{enumerate}

Transition system specifications allow
for the  formal description of overall  system
behavior in a syntax-oriented, compositional
way, using inference rules for local transitions and their
composition.  Process synchronization can be  expressed
abstractly using, e.g., conditions on system states and reachability
conditions over transition relations as premises, and label
synchronization for parallel transitions. This high level of
abstraction greatly simplifies the verification of system properties.
Whereas TSS is well-known as a formalism to define language semantics
and reason about language meta-theory, it is also well-suited to
describe specific systems in order to reason about, e.g., reachability
or state invariance.

For the second step, we need an implementation language with a formal
semantics (e.g., formalized by a TSS) which enables a simulation
relation to be formally established.
In this paper, we have opted for the 
\emph{active object} language ABS  \cite{ABS:tutorial,RTABS:tutorial} (ABS stands  for
\emph{Abstract Behavioral Specification}).
The semantics of the ABS language  is formally defined by a TSS~\cite{johnsen10fmco} and implemented by  backends\footnote{\url{https://abs-models.org/}} in Erlang, Haskell, and Java, all of  which support parallel
execution.
It has been developed and applied in the context of  various EU projects, e.g.,  in the EU FP7  projects
HATS\footnote{\url{https://cordis.europa.eu/project/id/231620}}
(\emph{Highly Adaptable and Trustworthy Software using Formal Models})
and ENVISAGE\footnote{\url{https://cordis.europa.eu/project/id/610582}} (\emph{Engineering Virtualized Services}).
In these projects,  ABS has been extended and successfully applied to the formal  modelling and analysis of {software product families}~\cite{ABS-SPL} and software services deployed on the Cloud~\cite{johnsen15jlamp}.
The ABS tool suite\,\cite{DBLP:conf/esocc/GouwMNZ16,DBLP:conf/esocc/BezirgiannisBG17,fide,SACO,DECO,syco,din15cade,crowbar} has been further applied to  case studies, targeting cloud-based frameworks~\cite{turin20isola,lin16fase,DBLP:conf/isola/JohnsenLY16,hyvar_sc2_paper,AlbertBHJSTW14},  railway operations~\cite{formbarscp} and computational biology.\footnote{\url{https://www.compugene.tu-darmstadt.de}}

The parallel execution of active objects (see \cite{deboer17csur} for
a survey of active object languages) is a direct consequence of
decoupling method execution from method invocation by means of
\emph{asynchronous} method invocations.  The ABS language further
integrates a strict \emph{encapsulation} of the local state of an
active object with explicit language constructs for the
\emph{cooperative scheduling} of its method executions.  Since the ABS
language is tailored to the description of distributed systems, it
abstracts from the order in which method invocations are generated.

In the definition of the simulation relation, cooperative scheduling
allows the interleaving of methods in an active object to match the
granularity of the transition rules of the corresponding TSS.
Moreover, the parallel execution of active objects in ABS satisfies a
\emph{global confluence} property which allows to express
\emph{locally} the proof conditions of the simulation relation in a
syntax-directed manner, abstracting from the fine-grained interleaving
of the method executions.

As a proof-of-concept we introduce our method by its application to a
parallel simulator of multicore memory systems. These memory systems
generally use caches to avoid bottlenecks in data access from main
memory, but caches introduce data duplication and require protocols to
ensure coherence.  Although data duplication is usually not visible to
the programmers, the way a program interacts with these copies largely
affects performance by moving data around to maintain coherence.  To
develop, test and optimize software for multicore architectures, we
need correct, executable models of the underlying memory systems.  A
TSS of multicore memory systems with correctness proofs for cache
coherency has been described in \cite{bijo17facs,bijo19scp}, together
with a prototype implementation in the rewriting logic system
Maude~\cite{maudebook}. However, this fairly direct implementation of
the TSS is not well suited to simulate large systems.  Therefore we
introduce in this paper a parallel implementation based on the active
object model of the ABS language and apply our method to a proof of
its correctness.

This paper extends \cite{bezirgiannis19fase} which describes a first
version of the use case.  The extension consists of a formalization of
the novel idea of annotating ABS programs with TSS rules and the use
of a global confluence property of the ABS semantics in the formal
semantics (and verification) of these annotations.  Because of the
absence of this high-level specification of the simulation relation
between the ABS program and the TSS, the ABS implementation in
\cite{bezirgiannis19fase} has been developed largely independent of
the TSS, which considerably complicated the correctness proof.  In
constrast, the application of our new methodology lead to a major
refactoring of the ABS implementation described in
\cite{bezirgiannis19fase}, reflecting a correctness-by-design
development methodology.

\paragraph{Plan of the paper}

In the following section we introduce the main concepts of the ABS
language and in Section~ \ref{sec:methodology} the use of transition
rules as annotations of ABS programs. In Section~\ref{sec:cachememory}
we introduce the runtime syntax of the multicore TSS and in
Section~\ref{sec:model} we discuss its ABS implementation.
Section~\ref{sec: correctness} then introduces the correctness proof.
Related work is discussed in Section~\ref{sec:related} and in
Section~\ref{sec:conclusion} some general conclusions are drawn and
future work discussed.


\section{ABS: Actors with Cooperative Concurrency}
\label{sec:abs}
ABS is a modeling language for designing, verifying, and executing
concurrent software \cite{ABS:tutorial,RTABS:tutorial}. The core
language combines the syntax and object-oriented style of Java with
the Actor model of concurrency~\cite{hewitt_universal_1973}, resulting
in active objects which decouple communication and synchronization
using asynchronous method calls and cooperative scheduling
\cite{deboer17csur}.  Asynchronous method calls generate processes
(which execute the called methods) within the called (active) object
and do not impose any synchronization between caller and callee.
Instead, synchronization between different objects happens using
(implicit) futures, with which the caller and callee may synchronise
independently, at different times.  Synchronization between different
processes within an object is captured using cooperative scheduling. A
process allows another process to be scheduled by means of explicit
suspension points; rescheduling at the suspension point may depend on
the resolution of a future or on a Boolean conditional. This mechanism
allows the interleaving of different processes to be captured very
precisely in ABS.  

The imperative layer of synchronization and
communication is complemented by a functional layer, used for
computations over the internal data of objects.  The functional layer
combines parametric algebraic datatypes (ADTs) and a simple functional
language with case distinction and pattern matching. ABS includes a
library with predefined (Int, Bool, etc.)  and parametric datatypes
(lists, sets, maps, etc.) All other types and functions are
user-defined.

In the following, the basic ABS instructions used in this paper (and
shown in Figure~\ref{fig:ABS:instr}) are explained in terms of some
general synchronization patterns.

\begin{table}[t]
\small
\centering
\begin{tabular}{ll} 
\toprule
\textbf{Instruction} & \textbf{Meaning} \\
\midrule
\Abs{new} C & Creation of an instance of class C\\
\Abs{switch (e)}\{\Abs{p$_1$ => s$_1$}$\cdots$\Abs{p$_n$ => s$_n$}\} & Pattern matching \\
\Abs{ await b} & Suspension on  a Boolean condition \\
\Abs{ await e}!\Abs{m($e_1,\ldots,e_n$)} & Suspension on termination of a asynchronous call\\
\Abs{e}!\Abs{m($e_1,\ldots,e_n$)} & Non-blocking asynchronous call \\
\Abs{e.m($e_1,\ldots,e_n$)} & Blocking synchronous call \\
\Abs{m($e_1,\ldots,e_n$)} & Inlined (recursive)  self- call  \\
\bottomrule
\end{tabular}
\caption{\label{fig:ABS:instr}Basic ABS instructions used in this
  paper.  Here, \Abs{b} is a Boolean expression, \Abs{e} and
  \Abs{e}$_i$ denote expressions.}
\end{table}

\subsection{Synchronization Patterns}\label{sec:trans-patt}
We discuss encodings in ABS of a basic locking mechanism, atomic operations,
and a broadcast mechanism for global synchronization (using barriers).

\paragraph{Locks}
\begin{wrapfigure}{r}{0.32\textwidth} 
 \vspace{-12pt}
\begin{absexamplen}{numbers=none}
class Lock {
  Bool unlocked = True;
  
  Unit take_lock{
    await unlocked;
    unlocked = False;
  }

  Unit release_lock{
    unlocked = True;
  }
}
\end{absexamplen}
\vspace{-10pt}
\caption{\label{fig:buslock} Lock implementation in ABS using
  await on Booleans.}
  \vspace{-10pt}
\end{wrapfigure} 
The basic mechanisms of asynchronous method calls and cooperative
scheduling in ABS can be explained by the simple code example of a
class \Abs{Lock} (Figure~\ref{fig:buslock}).  It uses an \Abs{await}
statement on a Boolean condition to model a binary semaphore, which
enforces exclusive access to a common resource ``lock'',
modeled as an instance of the class \Abs{Lock}
(dynamically  created by the execution of the expression \Abs{new}
\Abs{Lock}).  More
specifically, execution of the \Abs{take\_lock}
method  will be suspended by the \Abs{await unlocked} statement.  This
statement \emph{releases the control}, allowing the scheduling of
other (enabled) processes within the \Abs{Lock} object.  When the
local condition \Abs{unlocked} inside the \Abs{Lock} object has become
true, the generated \Abs{take\_lock} processes within the \Abs{Lock}
object will compete for execution.  The scheduled process then will
terminate and return by setting \Abs{unlocked} to \Abs{False}.

In general, the \emph{suspension points} defined by
\Abs{await} statements define the granularity of interleaving of the
processes of an object. The statement \Abs{await
  lock}!\Abs{take_lock()} will only suspend the process that issued the call (and release control in the caller object) until
\Abs{take\_lock} has returned.
In contrast, a \emph{synchronous} call \Abs{lock.take\_lock()} in ABS will generate a process for the execution of the \Abs{take\_lock()}  method
by the \Abs{lock} object and   block
(all the processes of) the caller object until the method returns. 

\begin{wrapfigure}[7]{r}{0.32\textwidth} 
\vspace{-26pt}
\begin{absexamplen}{numbers=none}
Bool TestandSet (/*input*/){ 
  Bool fail = False;
  switch /*test(input)*/$\;${ 
    True => /*set*/;
    False => fail = True;$\;$}
  return fail;$\;$}
\end{absexamplen}
\vspace{-10pt}
 \caption{\label{fig:TAS}Test and set pattern in ABS.}
\end{wrapfigure} 

 \paragraph{Atomic operations} 
 The interleaving model of concurrency of ABS allows for a simple and
 high-level implementation of atomic operations.  For example,
 Figure~\ref{fig:TAS} shows a general ABS implementation of
 test-and-set instructions~\cite{andrewsParallelProg}, where the
 concurrency model guarantees that the local \Abs{/*test(input)*/} and
 \Abs{/*set*/} instructions, assuming that they do not involve
 suspension points, are not interleaved and thus can be thought of as
 executed in a single atomic operation.  In ABS test instructions can
 be implemented using the \Abs{switch}-instruction, which evaluates an
 expression that matches the resulting value against a pattern
 \Abs{p} in the different branches.  This instruction has been mainly
 used to pattern match the ADTs used in the ABS program
discussed in this paper. In the
 simplest case, this pattern can be replaced by an
 \Abs{if}-\Abs{else}-instruction.
 Instances of this atomic pattern can be observed in
 Figures~\ref{fig:abs-aux}~and~\ref{fig:abs-swap}, in the methods
 \Abs{remove_inv} and \Abs{swap}.

\begin{figure}[h]
\centering
\begin{minipage}{0.45\textwidth}
\vspace{7pt}
\includegraphics[width=\textwidth]{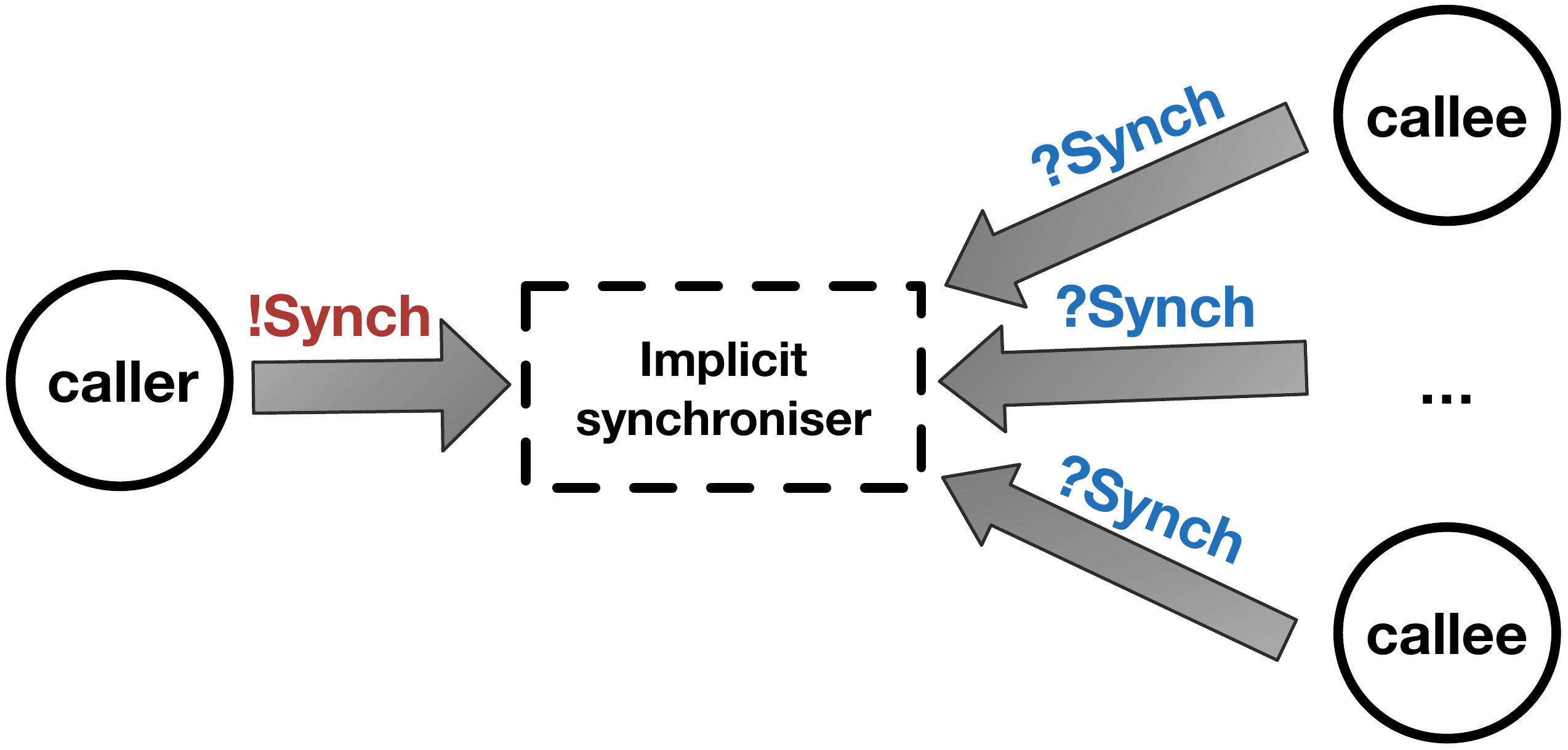}\\[16pt]
\begin{scriptsize}%
a) Broadcast synchronization in a labelled TSS.
\end{scriptsize}
\end{minipage}
\quad
\begin{minipage}{0.5\textwidth}
\includegraphics[width=\textwidth]{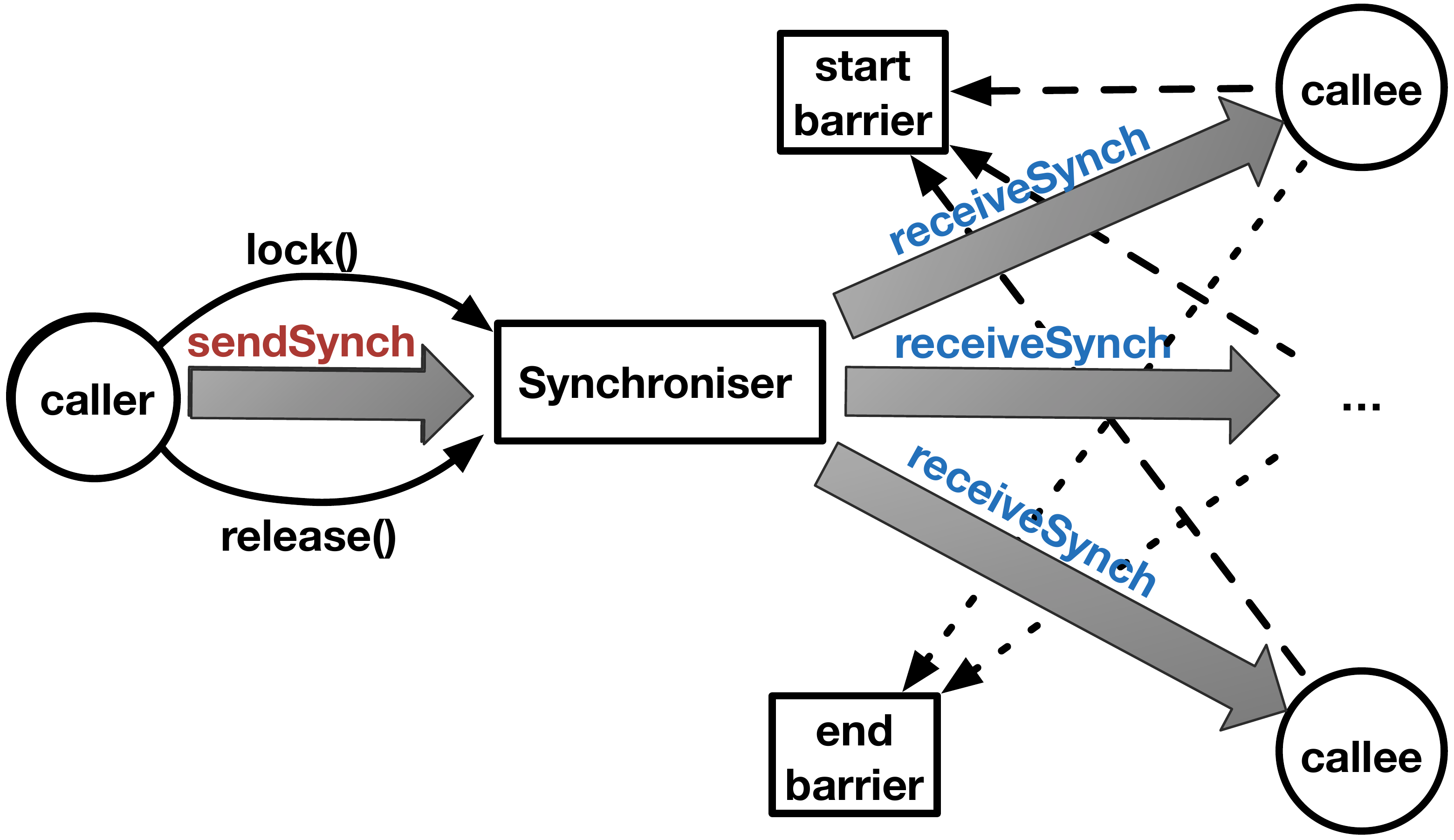}
\begin{scriptsize}%
b) Broadcast  synchronisation using an explicit synchroniser\\[-3pt]
and barriers in the ABS model.
\end{scriptsize} 
\end{minipage}
\caption{\label{fig:synccomp}Broadcast synchronisation patterns in TSS and ABS.}
\end{figure}

\paragraph{Broadcast synchronization}
Figure~\ref{fig:synccomp}a shows how broadcast synchronization in a
labelled TSS can be enforced simply by matching labels (an example is
detailed in Section~\ref{sec:cachememory}), thus abstracting from the
implementation details of the implicit multi-party synchroniser.  On the other hand, in programming languges like ABS the multi-party label synchronization needs to
be programmed explicitly; Figure~\ref{fig:synccomp}b illustrates the
architecture of  the ABS implementation 
in Figure~\ref{fig:absglobalsync}.

\begin{figure}[h!]
\begin{absexamplesm}
Interface IBroadcast { 
Bool broadcastSync(...);
Unit receiveSync (IBarrier start, IBarrier end, ...)}

Class Broadcast  implements IBroadcast, ...{
Bool broadcastSync(...){
   Bool signal=False;
   await sync!lock(); 
   if /*test*/ { sync.sendSync(this,...); /*set*/; signal=True; }
   sync.release();
   return signal
}

Unit receiveSync(IBarrier start, IBarrier end, ...) {
   $\label{lst:syncpatt.endsync2}$   await start!synchronise(); 
   /*some local computation*/; 
   end.synchronise(); }
...
}

Class Synchroniser (Set<IBroadcast> network) implements ISynchroniser {
Bool unlocked = True;
Unit lock(){ await unlocked; unlocked = False;$\;$}
Unit release(){unlocked = True;$\;$}
Unit sendSync(IBroadcast caller,...) {
     Set<IBroadcast> receivers = remove(network,caller);
     Int nrrecs= size(receivers);
     IBarrier start = new Barrier(nrrecs);
     IBarrier end = new Barrier(nrrecs+1); 
     foreach (receiver in receivers) { receiver!receiveSync(start,end,...); }
     end.synchronise();}
 } 


class Barrier(Int participants) implements IBarrier {
  Unit synchronise() { participants = participants - 1; await (participants == 0);$\;$}
}
\end{absexamplesm}
 \caption{Global synchronisation pattern in ABS.}
   \label{fig:absglobalsync}
\end{figure}
The class \Abs{Broadcast} serves as a template (or design pattern) for the implementation of a broadcast mechanism between its instances
which is specified by the interface \Abs{IBroadcast}.
The \Abs{broadcastSync} method encapsulates a synchronisation
protocol between \Abs{Broadcast} instances which uses  the additional classes \Abs{Synchroniser} and
\Abs{Barrier}.  This protocol consists of a synchronous call to the
method \Abs{sendSync} of an instance of the class
\Abs{Synchronise} (denoted by \Abs{sync})   which in turn asynchronously calls the method \Abs{receiveSync} of the
objects stored in the set \Abs{network} of  \Abs{Broadcast} instances,
excluding the  caller object executing the \Abs{broadcastSync} method.
We abstract from whether  the \Abs{sync} object
is passed as parameter of the \Abs{broadcastSync} method 
or part of the local state of any \Abs{Broadcast} instance.
The local computation specified by the \Abs{receiveSync} method   by the objects
in  \Abs{receivers} is synchronized by  calls of the method
\Abs{synchronise} of the new instances \Abs{start} and \Abs{end}
of class \Abs{Barrier}.
That is,   execution of this method by the  \Abs{start} and \Abs{end} barriers synchronise the start and the termination of the 
execution of the method \Abs{receiveSync}  by the objects
in  \Abs{receivers} and termination of the \Abs{sendSync} method itself.
This is achieved by a  ``countdown''  of the number of objects
in \Abs{receivers} that have called the \Abs{synchronise} method
plus one,  in case of  the \Abs{end} barrier.
The \Abs{synchronise} method of the
\Abs{start} barrier is called asynchronously
(Line~\ref{lst:syncpatt.endsync2}) and introduces a release point in
order to avoid a deadlock that may arise when an object that has not
yet called the \Abs{synchronize} method of the \Abs{start} barrier is
blocked on a synchronous method call to an object that has already
invoked (synchronously) the \Abs{synchronize} method of the
\Abs{start} barrier. On the other hand, the corresponding call to the
\Abs{end} barrier is synchronous to ensure that  all the objects
in \Abs{receivers} have completed their
local computations.  
The additional synchronisation of  the synchroniser object on the \Abs{end} barrier ensures that also the caller of the \Abs{sendSync} method is
blocked until all the local computations specified by the  \Abs{receiveSync} method have been completed.

Objects in ABS are input enabled, so it is always possible to call a
method on an object.  In our implementation, this scheme could give
rise to inconsistent states if several objects start the protocol in
parallel.
To ensure exclusive access to the synchroniser at the start of the
protocol, we add a lock to the synchroniser protocol, such that the
caller must take the lock before calling \Abs{sendSync} and release
the lock upon completion of the call. The resulting exclusive access
to the synchroniser guarantees that its message pool contains at most
one call to the method \Abs{sendSync}.

\subsection{Semantics}
ABS is a formally defined language \cite{johnsen10fmco}; in fact, its (operational)
semantics is defined by a TSS which allows us to reason formally about
the execution of ABS programs.  The semantics of an ABS model can be
described by a transition relation between global configurations. A
global configuration is a (finite) set of object configurations.  An
object configuration is a tuple of the form
$\langle \oid, \sigma, p, Q\rangle$, where $\oid$ denotes the unique
identity of the object, $\sigma$ assigns values to the instance
variables (fields) of the object, $p$ denotes the currently executing
process, and $Q$ denotes a set of (suspended) processes (the object's
``queue''). A process is a closure $(\tau,S)$ consisting of an
assignment $\tau$ of values to the local variables of the statement
$S$.  We refer to \cite{johnsen10fmco} for the details of the TSS for
deriving transitions $G\rightarrow G'$ between global configurations
in ABS.

Although only one thread of control can
execute in an active object at any time, cooperative scheduling allows
different threads to interleave at explicitly declared points in the
code, i.e., the \Abs{await} statements.
When the currently executing process is suspended by an \Abs{await}
statement, another (enabled) process is scheduled. Access to an
object's fields is protected: any non-local (outside of the object)
read or write to fields happens explicitly via method calls so as to
mitigate race-conditions or the need for extensive use of explicit
mutual exclusion mechanisms (locks).

Since active objects only interact via method calls and processes are
scheduled non-deterministically, which provides an abstraction from
the order in which the processes are generated by method calls, the
ABS semantics satisfies the following global confluence property
(see also \cite{bezirgiannis19fase,tveito20fase})  that
allows to commute consecutive local computations steps of processes
which belong to \emph{different} objects.

\begin{theorem}[Global confluence]
  \label{lemma:global.confluence}
  For any two transitions $G_1\rightarrow G_2$ and $G_1\rightarrow G_3$
  that describe execution steps of  processes of different
  objects, there exists a global configuration $G_4$ such that
  $G_2\rightarrow G_4$ and $G_3\rightarrow G_4$.
\end{theorem}

An important consequence of the above global confluence property,
which underlies the main results of this paper, is that we can
restrict  the global
interleaving between processes  by  reordering
the execution steps in an ABS computation.
In particular,
we can restrict the interleaving
semantics of the ABS model taking into account general semantic
properties of  synchronous communication, 
and the implementation of locks
and broadcast synchronization in ABS, as explained next.

Since a synchronous call of a method of  \emph{another} object in ABS,  blocks
all processes of the caller (object), the global confluence property
allows further restricting 
the interleaving of the ABS
processes
so that the caller process is resumed
\emph{immediately} after the synchronous method invocation has
terminated.

It is worthwhile to note that in general we can \emph{not} assume that
a method that is called synchronously in ABS is also scheduled
\emph{immediately for execution} because this would discard execution
of other processes by the callee.  

The global confluence property also allows abstracting from the
internal computation steps of the above ABS implementation of the
global (broadcast) synchronization pattern because it allows
scheduling the processes generated by the \emph{broadcast} method so
that execution of this method  is not interleaved with any other processes.

We can formalize the above in terms of the following notion
of \emph{stable} configurations.

\begin{definition}[Stable configurations]
  An object configuration is {stable} if the statement to be executed
  denotes the termination of an \emph{ asynchronously} called method
  (we assume a special runtime syntax which denotes such termination),
  or it starts with a synchronous call \emph{to another object} or a
  \Abs{await} statement.  A global ABS configuration is {stable} if
  all its object configurations are stable.
\end{definition}

Note that since synchronous self-calls are executed by inlining they
do not represent an interleaving point.

In the sequel $G \Rightarrow G'$ denotes the transition relation which
describes execution starting from a global stable configuration $G$ to
a next one $G'$ (without passing intermediate global stable
configurations).  We distinguish the following three
cases:
\begin{enumerate}
\item
The transition $G \Rightarrow G'$ describes the \emph{local} execution of a method by a single object.
\item 
The transition $G \Rightarrow G'$ describes the \emph{rendez-vous} between
the caller and callee of a synchronous method  call in terms of the terminating execution of the  called method, \emph{followed} by
the resumption of the suspended call.
\item
The transition $G \Rightarrow G'$ describes the effect of executing the \Abs{broadcast} method, which thus describes the \emph{global} synchronization of different objects.
\end{enumerate}

This coarse-grained interleaving
semantics of ABS forms the basis for the general methodology to prove
correctness of ABS implementations of TSS specifications, described
next.


\section{The General Methodology}\label{sec:methodology}
\subsection{Annotating ABS with TSS Rules}\label{sec:Anno}
For a general introduction of transition system specifications  we  refer to \cite{G19}.
The general methodology for the development of ABS implementations of
abstract TSSs is based on the coarse-grained interleaving described in
Section~\ref{sec:abs} (denoted by the transition relation
$\Rightarrow$): it allows focusing on the design of \emph{local,
  sequential} code that implements the individual transition rules.
This is reflected by the following use of transition rules as a
\emph{specification formalism} of ABS code.  A \emph{conditional
  transition rule} $b:R$ consists of a local Boolean condition $b$ in
ABS and a name $R$ of a transition rule.  We use sequences
$b_1:R_1; \ldots; b_n:R_n$ of conditional transition rules to annotate
\emph{stable points}.  A stable point of a method definition denotes
either its body or a substatement of its body that starts with an
external synchronous call or an \Abs{await} statement.  The idea is
that each $b_i$ is evaluated as a condition which identifies a
\emph{path} leading from the annotated stable point to a next one or
to termination.  The execution of this path should correspond to the
application of the associated transition rule $R_i$. This
correspondence involves a simulation relation, described below.

A sequence $b_1:R_1; \ldots; b_n : R_n$ of conditional transition
rules is evaluated from left to right, that is, the first transition
rule from the left, the Boolean condition of which evaluates to true,
is applicable.  The case that all Boolean conditions are false means
that there does not exist a transition rule for \emph{any} path from
the annotated stable point to a next one or to termination (in the simulation relation
all these paths would correspond to a ``silent'' transition).  As a
special case, we stipulate that for \emph{any} path leading from a
stable point \emph{which has no associated annotation} to a next
stable point (or to termination)  there does \emph{not} exist a corresponding transition
rule. The use of annotations in the ABS code of the multicore memory
system is shown in Section~\ref{sec:behv-view}.

\subsection{Correctness of the Implementation}\label{sec:method}

The correctness of the ABS implementation with respect to a given
TSS can be established by means of a simulation relation between the
transition system describing the semantics of the ABS implementation
and the transition
system describing the TSS.
The annotation of ABS code with (conditional) TSS rules provides a
high-level description of the simulation relation, describing which
rule(s) correspond with the execution of the ABS code from one stable
point to a next one (or to termination).  Underlying this high-level description, we define a simulation relation between ABS configurations and the runtime
states of the 
TSS.
This  simulation
relation  is defined as an abstraction function $\alpha$ which maps every stable
global ABS configuration $G$ to a behavioral  equivalent
TSS configuration $\alpha(G)$ (see Section  \ref{sec: correctness}).

We restrict the simulation relation to \emph{reachable} ABS
configurations.  A configuration $G$ of the ABS 
program is {reachable} if $G_0\Rightarrow^* G$, for some
\emph{initial} configuration $G_0$. 
  In an initial configuration of
 the ABS multicore program all process queues are empty, and the only
 active processes are those about to execute the run methods of the
 cores.  
This restriction allows to use some general properties of the
ABS semantics; e.g., upon return of a synchronous call, the local
state of the calling object has not changed.

We can now express that a  ABS 
program
is a correct implementation of a 
TSS specification by proving that the following
theorem holds, given an abstraction function $\alpha$:

\begin{definition}[Correctness]
  \label{theorem:simulation-relation-methodology}
  Given an ABS program and a TSS, let $\alpha$ be an
  abstraction function from configurations of the ABS program to TSS configurations. The ABS program is a correct implementation of
the TSS,
  if for any reachable configuration $G$ and transition
  $G\Rightarrow G'$ of the ABS program we have that $\alpha(G)=\alpha(G')$ or
  $\alpha(G)\rightarrow\alpha(G')$.
\end{definition}

Because of the general
confluence property of the ABS semantics to prove that $\alpha$ is a simulation relation, it suffices to verify the
annotations of 
methods 
in terms of the abstraction function  $\alpha$.  The general idea is
that for each transition $G\Rightarrow G'$ which results from the
execution from one stable point to a next one (or to termination), we have to show that
$\alpha(G')$ results from $\alpha(G)$ by application of the enabled
TSS rule associated with the initial stable point. In case no TSS rule
is enabled, we have a ``silent'' step, that is,
$\alpha(G)=\alpha(G')$.


\section{A TSS for Multicore Memory Systems}
\label{sec:cachememory}
\begin{wrapfigure}{r}{0.5\textwidth} 
\vspace{-12pt}
  \fbox{\includegraphics[width=0.97\linewidth]{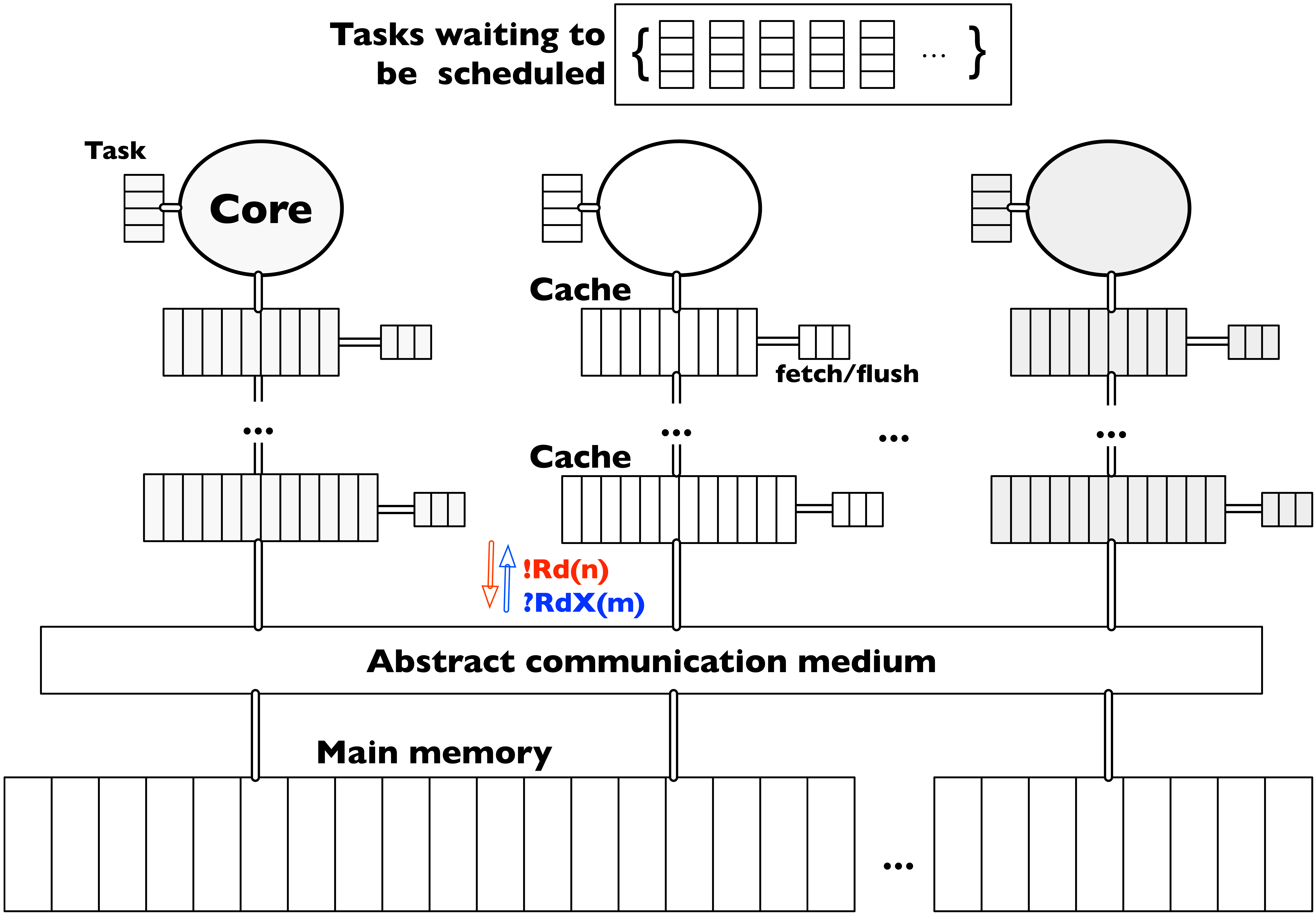}}
  \caption{ \label{fig:whitebox} Abstract model of a multicore memory system.}
\end{wrapfigure} 
Design decisions for programs running on top of a multicore memory
system can be explored using simulators (e.g.,
\cite{binkert11can,carlson.heirman.eeckhout:sniper,martin05can,miller*:graphite}).
Bijo et al.\ developed a TSS for multicore memory systems
\cite{bijo17facs,bijo19scp}.  Taking this TSS as a starting point, we
will study how a parallel simulator can be developed which implements
the TSS and use this development to discuss the details of our
methodology. We first introduce the main concepts of multicore memory
systems and then look at their formalization in terms of a TSS.

A multicore memory system consists of cores that contain \emph{tasks}
to be executed, the \emph{data layout} in main memory (indicating
where data is allocated), and a system \emph{architecture} consisting
of cores with private multi-level caches and shared memory (see
Figure~\ref{fig:whitebox}).  Such a system is parametric in the number
of cores, the number and size of caches, and the associativity and
replacement policy.  Data is organized in blocks that move between the
caches and the main memory.  For simplicity, we abstract from the data
content of the memory blocks, assume that the size of cache lines and
memory blocks in main memory coincide and that a local reference to a
memory block is represented directly by the corresponding memory
address, and transfer memory blocks from the caches of one core to the
caches of another core via the main memory. As a consequence, the
tasks executed in the cores are represented as data access patterns,
abstracting from their computational content.
 
Task execution on a core requires memory blocks to be transferred from
the main memory to the closest cache.  Each cache has a pool of
instructions to move memory blocks among caches and between caches and
main memory. Memory blocks may exist in multiple copies in the memory
system.  Consistency between different copies of a memory block is
ensured using the standard cache coherence protocol MSI
(e.g.,~\cite{solihin2015fundamentals}), with which a cache line can be
either modified, shared or invalid. A \emph{modified} cache line has
the most recent value of the memory block, therefore all other copies
are \emph{invalid} (including the one in main memory). A \emph{shared}
cache line indicates that all copies of the block are consistent.  The
protocol's messages are broadcasted to the cores. The details of the
broadcast (e.g., on a mesh or a ring) can be abstracted into an
\emph{abstract communication medium}.  Following standard
nomenclature, \emph{Rd} messages request \emph{read} access and
\emph{RdX} messages \emph{read exclusive} access to a memory
block. The latter invalidates other copies of the same block in other
caches to provide write access.

We summarize the operational aspects of cache coherency with the MSI
protocol.  To access data from a memory block~$n$, a core looks
for~$n$ in its local caches.  If~$n$ is not found in shared or
modified state, a \emph{read request} $\sendR{n}$ is broadcasted to the
other cores and to main memory. The cache can \emph{fetch} the block
when it is available in main memory. Eviction is required if the cache
is full, removing another memory block to free space. Writing to
block~$n$ requires $n$ to be in shared or modified state in the local
cache; if it is in shared state, an \emph{invalidation request}
$\sendRX{n}$ is broadcasted to obtain exclusive access.  If a cache with
block $n$ in modified state receives a read request~$\recR{n}$, it
\emph{flushes} the block to main memory; if a cache with block $n$ in
shared state receives an invalidation request~$\recRX{n}$, the cache
line will be \emph{invalidated}; the requests are discarded
otherwise. Read and invalidation requests are broadcasted
instantaneously in the abstract model, reflecting that signalling on
the communication medium is orders of magnitude faster than moving
data to or from main memory.

\begin{figure}[t!]
\small \centering
$\begin{array}{@{}l@{\hspace{24pt}}l@{}}
\begin{array}[t]{@{}l@{}}
\emph{\textbf{Syntactic}}\\\emph{\textbf{categories.}}\\
\cid  \in  \mathit{CoreId} \\
 \levid  \in  \mathit{CacheId}\\
n \in \mathit{Address}\\
\end{array}
&
\begin{array}[t]{@{}r@{\hspace{1pt}}c@{\hspace{2pt}}l@{\hspace{2pt}}c@{\hspace{2pt}}l@{}}
    \\
    \multicolumn{3}{@{}l@{}}{\emph{\textbf{Definitions.}}}\\
     \mathit{cf} & \in &  \config & ::= &  
\langle \many \core , \many \cache , \memory \rangle  \hist{: H} \\ 
    \core & \in &    \mathit{Core} & ::= &  \cid  \lsep \task \hist{: h} \\ 
    \cache &  \in &  \mathit{Cache} & ::=& \levid \lsep \memory \lsep  \datainst \\  
    \status &  \in  &    \mathit{Status} & ::= & \{ \mo, \sh , \inv\} \\ 
  \stask &\in &\mathit{AccessPtns} & ::=&  \none  \mid \stask; \stask \mid  \reads{n} \mid  \writes{n}    \\
   \task&\in & \mathit{RunLang}& ::= & \stask \mid \task; \task \mid \readBs{n} \mid \writeBs{n}  \\
    \datainst&\in& \mathit{DataLang}& ::= & \none \mid  
                                            \datainst \msunion \datainst\mid \fetchs{n}
                                            \mid \blocks{n}  \\
&&&&                                   \mid \fetchwait{n}{n'} \mid \flushs{n}
  \end{array}
\end{array} $
\caption{\label{fig:rts} Syntax of runtime configurations,
where over-bar denotes sets
    (e.g., $\many\core$).}
\end{figure}

\subsection{A TSS  of  Multicore Memory Systems\label{sec.formalization}}
The multicore TSS describes the interactions between a core, caches,
and the main memory.  It further includes labeled transitions to model
instantaneous broadcast.  In general a model of the multicore TSS is a
\emph{transition system}.  We refer to a model of the multicore TSS,
which is parametric in the number of cores and caches, also as a
Multicore Memory System (MMS, for short).  The multicore
TSS~\cite{bijo17facs,bijo19scp} is shown to guarantee correctness
properties for data consistency and cache coherence (see,
e.g.,~\cite{CullerBook,Sorin:2011}), including the preservation of
program order in each core, the absence of data races, and that stale
data is never accessed.

We outline the main aspects of a simplified version of the multicore
TSS which allows focusing on the main challenges of a correct
distributed implementation.  The runtime syntax is given in
Figure~\ref{fig:rts}.  A configuration $\mathit{cf}$ is a tuple
consisting of a main memory $\memory$, cores $\many\core$, caches
$\many\cache$ (we abstract from the tasks to be scheduled).
A core~$(\cid \lsep \task)$ with identifier~$\cid$ executes \emph{runtime
statements}~$\task$.  A cache $(\levid \lsep \memory \lsep \datainst )$
with identifier~$\levid$ has a local cache memory~$\memory$ and data
instructions~$\datainst$. We assume that the cache identifier $\levid$
encodes the $\cid$ of the core to which the cache belongs and its
level in the cache hierarchy.  We use $\mathit{Status}_{\bot}$ to
denote the extension of the set $\{ \mo, \sh , \inv \}$ of status tags
with the undefined value $\bot$. Thus, a memory
$\memory: \mathit{Address} \rightarrow \mathit{Status}_{\bot}$ maps
addresses~$n$ to either a status tag $\status$ or to $\bot$ if the
memory block with address~$n$ is not found in $\memory$.

\emph{Data access patterns} $\stask$ model tasks consisting of finite
sequences of $\reads{n}$ and $\writes{n}$ operations to address $n$
(that is, we abstract from control flow operations for sequential
composition, non-deterministic choice, repetition, and task creation).
The empty access pattern is denoted $\none$.  Cores execute runtime
statements $\task$, which extend~$\stask$ with $\readBs{n}$ and
$\writeBs{n}$ to block execution while waiting for data.  Caches
execute \emph{data instructions} from a multiset $\datainst$ to fetch
or flush a memory block with address~$n$; here, $\fetchs{n}$ fetches a
memory block with address $n$, $\blocks{n}$ blocks execution while
waiting for data, $\fetchwait{n}{n'}$ waits for a memory block $n'$ to
be flushed before fetching $n$ (this is needed when the cache is
full), and $\flushs{n} $ flushes a memory block.

The connection between the main memory and the caches of the different
cores is modelled by an \emph{abstract communication medium} which
allows messages from one cache to be transmitted to the other caches
and to main memory in a parallel instantaneous broadcast.
Communication in the abstract communication medium is captured in the
TSS by label matching on transitions.
For any address $n$, an output of the form $ \sendR{n}$ or
$\sendRX{n}$ is broadcasted and matched by its dual of the form
$\recR{n}$ or $\recRX{n}$.  The syntax of the model is further
detailed in~\cite{bijo17facs,bijo19scp}.  For a complete overview of
the transition rules we refer to \ref{sec:TSS}.
In the next section, we will introduce these
rules incrementally when discussing their ABS implementation.

The following auxiliary functions are used in the transition
  rules, given a cache identifier~$\levid$:
\begin{itemize}
\item $\coreid{\levid}$ returns the identifier of the core to which
  the cache belongs;
\item $\levno{\levid}$ gives the level at which the cache is located
  in the memory hierarchy;
\item $\first{\levid}$ is true when $\levno{\levid} =1$, otherwise
  false;
\item $\last{\levid}$ is true when $\levno{\levid} =l$ where $l$ is
  the number of caches in the hierarchy, otherwise false;
\item $\statusfunc{M}{n}$ returns the status of block $n$ in
  memory~$M$ or $\bot$ if the block is not found in~$M$; and
\item $\selectfunc{\memory}{n}$ determines the address where a
  block~$n$ should be placed in the cache, based on a cache
  associativity (e.g., random, set associativity or direct map) and a
  replacement policy (e.g., random or LRU).
\end{itemize}


\section{The ABS Model of the Multicore Memory System}
\label{sec:model}
This section describes the translation of the multicore TSS into a model in ABS 
\footnote{The ABS model for the multicore memory system can be found at \url{https://abs-models.org/documentation/examples/multicore_memory/}}.  
We
explain the structural and behavioural correspondence between these
two models.

\subsection{The Structural Correspondence}
\label{sec:struct-view}

\begin{figure}[t!]
\centering 
\vspace{-10pt}
\includegraphics[trim={10pt 10 10 11},clip, width=\textwidth]{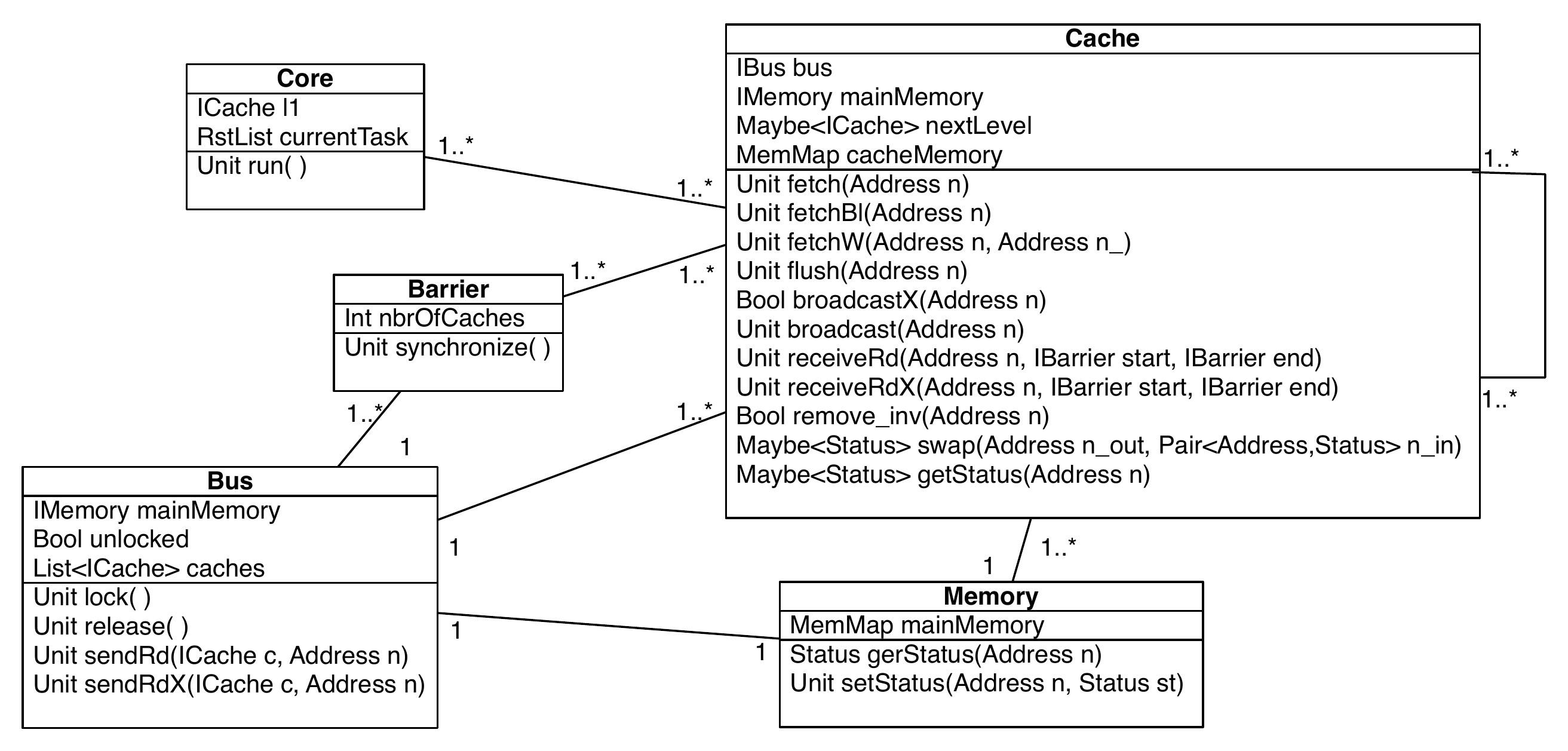}
\caption{ \label{fig:classDiagram} Class diagram of the ABS model.}
\bigskip
\bigskip
\includegraphics[trim={10pt 10 10 11},clip, width=.8\textwidth]{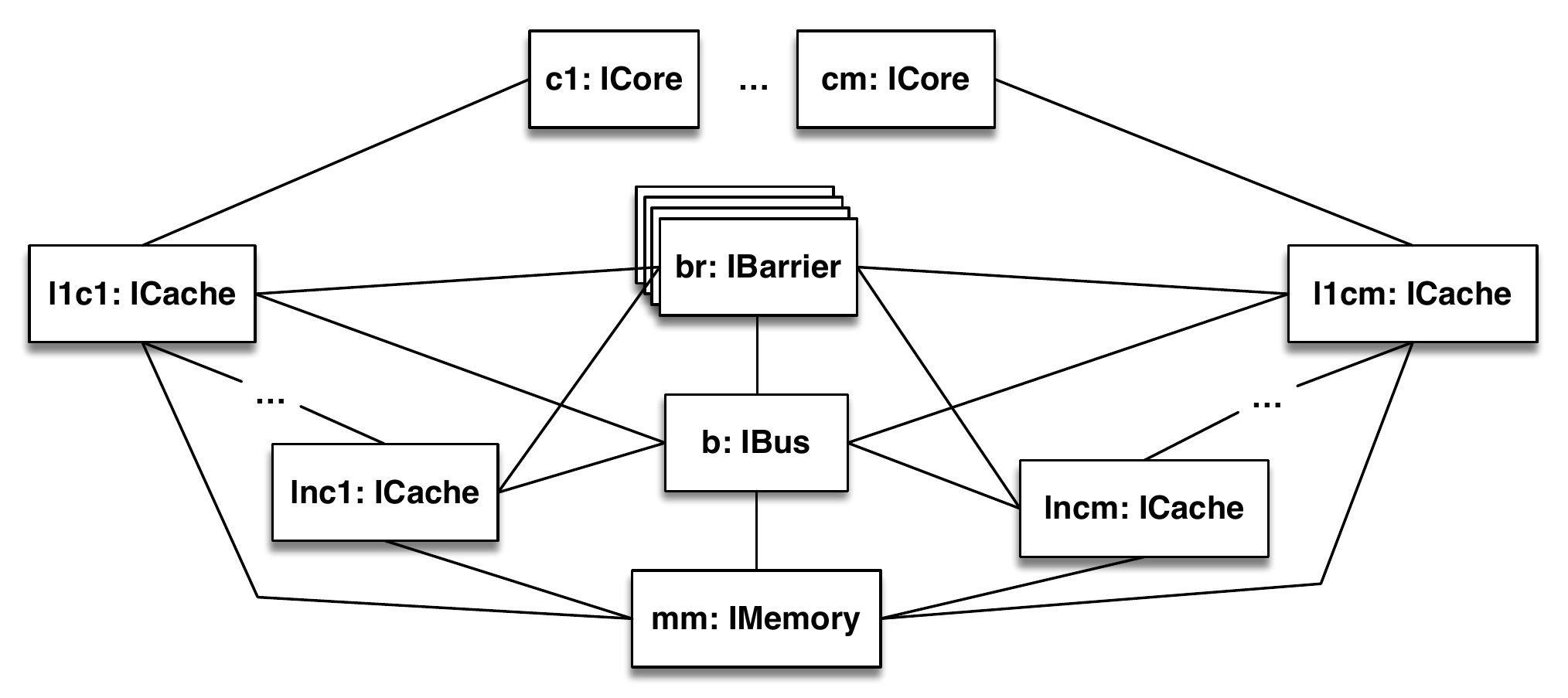}
\caption{ \label{fig:initialConfiguration} Object diagram of an initial configuration.}
\bigskip
\input{AnnotatedABS/ADT}
\caption{\label{fig:abs-ADT} Abstract data types of the model of the multicore memory system.}
\end{figure}

  The runtime syntax of the
multicore TSS is represented in ABS by classes, user-defined datatypes
and type synonyms, outlined in
Figures~\ref{fig:classDiagram}--\ref{fig:abs-ADT}.  An ABS
configuration consists of class  instances  to reflect the cores
with their corresponding cache hierarchies and the main memory.
Object identifiers guarantee unique names and object references are
used to capture how cores and caches are related. These references are
encoded in a one-to-one correspondence with the naming scheme of the
multicore TSS.

A core $\cid \lsep \task$ in the multicore TSS corresponds to an
instance of the class \Abs{Core} in ABS, where a field
\Abs{currentTask} of type \Abs{RstList} (as defined in
Figure~\ref{fig:abs-ADT}) represents the current list of runtime
statements . Each instance of the class \Abs{Core} further holds a
reference to the first level cache. 
An important design decision we made is to 
represent the  runtime statements $\task$ (of a core in  the multicore TSS)
as an ADT (see Figure ~\ref{fig:abs-ADT}).
A core in ABS then drives the simulation by processing these runtime statements which in general requires information about the first-level cache.
Alternatively, a core in ABS could delegate the processing of its runtime
statements by calling corresponding methods of the first-level cache.
However, this latter approach complicates the required callbacks.

A cache $\levid \lsep \memory \lsep \datainst$ in the multicore TSS
corresponds to an instance of class \Abs{Cache} with
a class parameter \Abs{nextLevel} which holds a reference to the next
level cache and a field \Abs{cacheMemory} which models the cache's
memory $\memory$ (of type \Abs{MemMap}, Figure~\ref{fig:abs-ADT}).
The multiset $\datainst$ of a cache's data instructions (see
Figure~\ref{fig:rts}) is represented by corresponding \emph{processes}
in the message pool of the cache object in ABS.  If the value of
\Abs{nextLevel} is \Abs{Nothing}, then the object represents the last
level cache (in the multicore TSS, a predicate $\flag$ is used to
identify the last level). 

In addition, the ABS implementation of the global
synchronisation with labels $\sendR{n}$ and $\sendRX{n}$ used in  the
multicore TSS is based on the global synchronisation pattern 
as described in Figure~\ref{fig:absglobalsync}.
However,
instead of distinguishing between these two labels by means of an additional parameter, we introduce two corresponding
broadcast interfaces:
¨

\begin{absexamplesm}
Interface IBroadcast { 
Bool broadcast(...);
Unit receiveRd (IBarrier start, IBarrier end, ...)}

Interface IBroadcastX { 
Bool broadcastX(...);
Unit receiveRdX (IBarrier start, IBarrier end, ...)}
\end{absexamplesm}

The class \Abs{Cache} then provides an implementation of both interfaces
following the template of the class \Abs{Broadcast} in Figure~\ref{fig:absglobalsync}.
The ABS class \Abs{Bus},  on the other hand,  follows the template
of the \Abs{Synchroniser} class with the two versions
\Abs{sendRd} and \Abs{sendRdX} of the method \Abs{sendSync}.

The object diagram in Figure~\ref{fig:initialConfiguration} shows an
initial configuration corresponding to the one depicted in
Figure~\ref{fig:whitebox}.

\subsection{The Behavioural Correspondence}
\label{sec:behv-view}
We next discuss the ABS implementation of the  transition rules
of the multicore TSS, and the ABS synchronization patterns
described in Section~\ref{sec:abs}. We
observe that the combination of \emph{asynchronous method calls} and
\emph{cooperative scheduling} in ABS is crucial because of the
\emph{interleaving} inherent in the multicore TSS, which requires that
objects are able to process other requests while executing a method in
a controlled way; e.g., caches need to flush memory blocks while
waiting for a fetch to succeed.

\subsubsection{The Annotated ABS Multicore Implementation}
\label{sec:behv-trans}
The classes \Abs{Core} and \Abs{Cache} pose the main implementation
challenges.  Here we explain the implementation of the \Abs{run}
method (Figure~\ref{fig:ABScore}) of the class \Abs{Core} (which is
its only method) informally, in terms of its annotations (as
introduced in Section~\ref{sec:Anno}).
In Section~\ref{sec: correctness} we introduce a formal semantics
of these annotations as a high-level description of a simulation relation,
and prove the correctness of the class \Abs{Cache}.

The \Abs{run} method may generate {synchronous} calls to the auxiliary
methods in Figure~\ref{fig:abs-aux}.  
The method \Abs{remove\_inv} instantiates the test-and-set pattern
of Figure \ref{fig:TAS}.
The method \Abs{broadcastX} 
is  an instance of the global synchronization pattern
described in Section~\ref{sec:abs},  Figure \ref{fig:absglobalsync}.   The method \Abs{sendRdX} of the
global synchroniser \Abs{bus} asynchronously calls the method
\Abs{receiveRdX}, see Figure~\ref{fig:abs_receiveRdx}, of all caches
(except for the calling cache), using the barrier synchronization
described in Section~\ref{sec:abs}.

\begin{figure}[t]
  \input{AnnotatedABS/runcore_abs}
  \caption{\label{fig:ABScore} Annotated \Abs{run} method.}
\end{figure}

\begin{figure}[h!]
\input{AnnotatedABS/ABSAux}
\caption{\label{fig:abs-aux}Methods \Abs{getStatus}, \Abs{remove_inv},   and \Abs{broadcastX} of class \Abs{Cache}.}
\bigskip
\input{AnnotatedABS/receiveRdx_abs}
\caption{\label{fig:abs_receiveRdx}  Annotated \Abs{receiveRdX}  method.}
\end{figure}

Since the stable point at the beginning of the \Abs{run} method has
no associated annotation, by definition (see Section~\ref{sec:Anno}),
for \emph{any} path from the beginning to a next stable point (or to termination) there
does \emph{not} correspond a transition rule (of the multicore TSS).
For example, there is no transition rule corresponding to the case
that the run method terminates when \Abs{curentTask==Nil} (note that
because of the structural correspondence also the corresponding core
has no runtime statements \emph{rst} to execute).  Similarly, there
are no transition rules corresponding to the execution of the code
from the beginning of the method to the synchronous calls to the
auxiliary methods~\Abs{remove\_inv} (Line~\ref{lst:run.removeRead})
and \Abs{getStatus} (Lines~\ref{lst:run.useReadl},
\ref{lst:run.useWrite}, \ref{lst:run.WriteBl.use}) of the first level
cache which, besides the pattern matching, only consists of the call
itself.

The condition of the annotation \Abs{removed==True} :
$\crule{\rn{PrRd}_2}$ (Line~\ref{lst:run.removeRead}) associated with
the synchronous call to the \Abs{remove\_inv} method describes the
path which leads from its execution and return via the
\Abs{then}-branch of the subsequent \Abs{if}-statement to the
termination of the run method (after it has called itself again
asynchronously).  According to the annotation, the execution of this
path corresponds to the $\crule{\rn{PrRd}_2}$ transition rule:
$$
 \ncondrule{PrRd$_2$}{
           \quad  \first{\levid} = true   \quad \coreid{\levid} = c 
        \quad
        \statusfunc{\memory}{n} \in \{\inv,\bot\} 
        }{
       (c  \lsep \hlc{\reads{n};\task}) \hsep   (\levid \lsep \hlcc{\memory \lsep \datainst})\hist{: h}
        \to \\
        (c  \lsep \hlc{\readBs{n};\task}) \hsep (\levid \lsep \hlcc{\memory  \setto{n}{\bot} \lsep  \datainst\msunion\fetchs{n}})  \hist{: h}
        }
$$
This rule handles the case when a core intends to read a memory block
with address~$n$, which is not found in the first level cache.  The
core will then be blocked while waiting for the memory block to be
fetched either from the lower level caches or main memory.  Note that
the condition as returned by the \Abs{remove\_inv} method signals that
the status of the address of the first level cache is undefined or
invalid.

On the other hand, the condition \Abs{removed==False} describes the
path which leads from its execution and return  via the \Abs{else}-branch (Line~\ref{lst:run.PrRd1}), which
also leads to the termination of this invocation of the  run method.  According to the
annotation, the execution of this path corresponds to the
$\crule{\rn{PrRd}_1}$ transition rule:
$$
\ncondrule{PrRd$_1$}{
        \quad   \first{\levid} = true   \quad \coreid{\levid} = c 
        \quad
        \statusfunc{\memory}{n} \in \{\sh,\mo\} 
        }{
       (c  \lsep \hlc{\reads{n};\task}) \hsep (\levid  \lsep \memory \lsep \datainst) \hist{: h}
        \to
      (c  \lsep \hlc{\task})  \hsep (\levid \lsep \memory \lsep \datainst) \hist{: h; R(c, n)} }
$$
This  rule covers the case when the memory block to be read by a core is
found in its first level cache. . Note that the condition as returned by
the \Abs{remove\_inv} method implies that the status of the address of
the first level cache 
is either shared or modified.

Next we consider the annotation \Abs{status} != \Abs{Nothing} :
$\crule{\rn{PrRd}_3}$ of the synchronous call to the \Abs{getStatus}
method (Line~\ref{lst:run.useReadl}).  Its condition describes the
execution path which leads from  the execution and return of the called
 \Abs{getStatus} method to termination of the \Abs{run} method via
the \Abs{then}-branch of the subsequent \Abs{if}-statement
(Line~\ref{lst:run.ifReadl}).  According to the annotation, the
execution of this path corresponds to the $\crule{\rn{PrRd}_3}$
transition rule:
$$
  \ncondrule{PrRd$_3$}{
          \first{\levid} = true \quad \coreid{\levid} = c 
          \quad
         n \in \dom{\memory}
        }{
         (c  \lsep \hlc{\readBs{n};\task}) \hsep  (\levid   \lsep \memory \lsep \datainst)\hist{: h}
          \to 
          (c  \lsep \hlc{\reads{n};\task}) \hsep  (\levid  \lsep \memory \lsep \datainst)  \hist{: h; R(c, n)}
        }
$$
This rule unblocks the core from waiting when~$n$ (i.e., the block to
be read) is found in the first level cache.
On the other hand, there does not exist a transition rule which
corresponds to the execution path described by the condition
\Abs{status==Nothing}.  This path leads from the execution of  the called \Abs{getStatus} method directly to the
termination of the \Abs{run} method without an update of the (local)
state, e.g., \Abs{currentTask} is not updated. In other words, the
evaluation of the \Abs{readBl(n)} instruction in ABS involves
\emph{busy waiting} until the status returned by the first level cache
is defined.  Alternatively, this could be implemented by calling
synchronously a method of the first level cache which simply executes
the statement \Abs{await lookup(cacheMemory,n)}!=\Abs{Nothing}.

The annotation of the synchronous call to the method
\Abs{getStatus} (Line~\ref{lst:run.WriteBl.use}) involves the rule
$$
          \ncondrule{PrWr$_4$}{
          \first{\levid} = true \quad \coreid{\levid} = c  
          \quad
         n \in \dom{\memory}
        }{
         (c  \lsep \hlc{\writeBs{n};\task})\hsep   (\levid   \lsep \memory \lsep \datainst) \hist{: h}
          \to
           (c  \lsep \hlc{\writes{n};\task}) \hsep (\levid  \lsep \memory \lsep \datainst) \hist{: h}
        }
 $$
 This annotation is explained in a similar manner as the annotation of
 the synchronous call to the \Abs{getStatus} method on
 Line~\ref{lst:run.useReadl}.  This rule unblocks the core from
 waiting when~$n$ (i.e., the block to be written) is found in the
 first level cache.

 We consider next the annotation \Abs{status}==J\Abs{ust(Mo)} :
 $\crule{\rn{PrWr}_1}$ of the synchronous call to the method
 \Abs{getStatus} (Line~\ref{lst:run.useWrite}) .  Its condition
 describes the execution path which leads from the execution  of the called \Abs{getStatus} method and subsequent
 execution of the \Abs{switch} statement to termination of the
 \Abs{run} method.  According to the annotation, the execution of this
 path corresponds to the $\crule{\rn{PrWr}_1}$ transition rule:
$$
   \ncondrule{PrWr$_1$}{
          \first{\levid} = true \quad \coreid{\levid} = c  
          \quad
          \statusfunc{\memory}{n}  = \mo
        }{
           (c  \lsep \hlc{\writes{n};\task}) \hsep  (\levid   \lsep \memory \lsep \datainst) \hist{: h}
          \to 
           (c  \lsep \hlc{\task}) \hsep  (\levid  \lsep \memory \lsep \datainst) \hist{: h; W(c, n)}
        }
$$
This rule allows a core to write to memory block~$n$ if the block is
found in a modified state in the first level cache.  On the other
hand, in case the condition does not hold, according to the annotation
no transition rules correspond to the execution paths which lead from
the execution of the called \Abs{getStatus} method to
the next stable points, i.e., the synchronous calls to the methods
\Abs{broadcastX} and \Abs{remove\_inv}
(Lines~\ref{lst:run.Write.sh.broadcastX} and \ref{lst:run.Write.else},
respectively).

The condition of the annotation \Abs{res==true} :
$\crule{\rn{PrWr}_2/\rn{SynchX}}$ of the synchronous call to the
\Abs{broadcastX} method (Line~\ref{lst:run.Write.sh.broadcastX}) of
the first level cache describes the path which leads from the  execution 
of the \Abs{broadcastX} method, followed by the execution of the subsequent
if-statement to termination of the run method (after an update of
\Abs{currentTask} and calling the \Abs{run} method again
asynchronously).  According to the annotation this path corresponds to
the global synchronization rule
$$
\ncondrule{SynchX}{
         
        \core\not\in \many{\core_1}\quad
        \core, \many{\cache}\hist{: H}
         \toL{\red{\sendRX{n}}} 
        \core', \many{\cache'}\hist{: H'} 

        }{
       \langle  \many{\core_1} \cup\{\core\} \hsep \many{\cache} \hsep \memory\hist{: H} \rangle 
        \to
      \langle   \many{\core_1}\cup\{\core'\}  \hsep \many{\cache'} \hsep  \memory\setto{n}{\pr{\_}{\inv}}  \hist{: H'}\rangle }
        $$
where the second premise is generated by successive  applications of the rule
$$
   \ncondrule{Synch-DistX}{
        \cache_1\not\in \many{\cache} \quad
        \core,\many{\cache} 
         \toL{\red{\sendRX{n}}} 
         \core', \many{\cache'} 
        \quad
        \cache_1
         \toL{\red{\recRX{n}}}
       \cache_2
        }{
        \core, \many{\cache}\cup \{\cache_1\}
        \toL{\red{\sendRX{n}}} 
         \core', \many{\cache'}\cup  \{\cache_2\}
        } 
        $$        
This latter rule itself is triggered by the following rules
$$
   \ncondrule{PrWr$_2$}{
         \first{\levid} = true   \quad \coreid{\levid} = c 
        \quad
         \statusfunc{\memory}{n} = \pr{k}{\sh}
        }{
        (c  \lsep \hlc{\writes{n};\task}) \hsep  (\levid   \lsep \hlcc{\memory} \lsep \datainst)\hist{: h}
        \toL{\red{\sendRX{n}}} 
         (c  \lsep \hlc{\task}) \hsep (\levid  \lsep \hlcc{\memory  \setto{n}{\pr{k}{\mo}}} \lsep \datainst) \hist{: h; W(c, n)}
        }        
$$

$$
  \ncondrule{Invalidate-One-Line}{
          \statusfunc{\memory}{n}= \sh
        }{
          \levid  \lsep \hlc{\memory} \lsep  \datainst 
          
        \toL{\red{\recRX{n}}}
        \levid  \lsep \hlc{\memory\setto{n}{\inv}} \lsep   \datainst   
      }
$$      
and
$$
  \ncondrule{Ignore-Invalidate-One-Line}{ 
       \statusfunc{\memory}{n} \in \{\inv,\bot\} 
      }{
        \levid  \lsep \memory \lsep  \datainst   
        
        \toL{\red{\recRX{n}}} 
        \levid  \lsep \memory \lsep  \datainst 
     }    
$$
These rules together capture the broadcast mechanism for invalidation
in the multicore memory system.  Rule \rn{PrWr$_2$}
corresponds to the case where a core writes to a memory block~$n$
that is marked as shared in its first level cache, which requires
broadcasting an invalidation message, $\sendRX{n}$,
to all the other caches.  This is achieved by triggering the global
synchronization rules \rn{SynchX} and \rn{Synch-DistX}.  While the
former identifies the core~$\core$
that broadcasts the invalidation message, the latter recursively
propagates the message, $\recRX{n}$,
to the other caches.  Depending on the local status of memory
block~$n$
in the recipient cache, the recipient cache will either invalidate the
local copy of the block (\rn{Invalidate-One-Line}), or ignore the
message (\rn{Ignore-Invalidate-One-Line}).

To explain this application of the $\crule{\rn{SynchX}}$
rule, we have a closer look at the definition of the \Abs{broadcastX}
method. Its body involves an instance of the global synchronization
pattern (Figure~\ref{fig:absglobalsync}).  As discussed in
Section~\ref{sec:abs}, because of the global confluence property, we
may assume that its execution is atomic, i.e., not interleaved with
any process that it has not generated.  The synchronous call to the
\Abs{sendRdX} method of the bus generates asynchronous calls to the
\Abs{receiveRdX} method (Figure~\ref{fig:abs_receiveRdx}) of all
caches except the one that initiated the global bus synchronization.
Following the general global synchronization pattern
(Figure~\ref{fig:absglobalsync}), these method calls are synchronized
by a \Abs{start} and an \Abs{end} barrier.  The two conditions of the
annotation at the beginning of the \Abs{receiveRdX} method describe
the two possible execution paths and their corresponding transition
rules $\crule{\rn{Invalidate-One-Line}}$
and $\crule{\rn{Ignore-Invalidate-One-Line}}$.

In case the condition \Abs{res==true} does not hold, according to the
annotation, no transition rule corresponds to the execution of the
\Abs{broadcastX} method. In this case the bus synchronization, as
invoked by the \Abs{broadcastX} method (Figure~\ref{fig:abs-aux}),
failed because the status of the address of the first level cache is
not shared anymore (as required by the $\crule{\rn{PrWr}_2}$
rule). Consequently, the processing of the \Abs{write(n)} instruction
itself fails and it will be processed again by the asynchronous self
call to the \Abs{run} method.

We conclude the informal explanation of the annotated \Abs{run} method
with the annotation\linebreak \Abs{removed==True}~:~$\crule{\rn{PrWr}_3}$ of the
synchronous call to the method \Abs{remove\_inv}
(Line~\ref{lst:run.Write.else}).  Its condition describes the path
that corresponds to the transition rule:
$$
   \ncondrule{PrWr$_3$}{
         \first{\levid} = true   \quad \coreid{\levid} = c 
        \quad
        \statusfunc{\memory}{n} \in \{\inv,\bot\}         }{
        (c  \lsep \hlc{\writes{n};\task}) \hsep (\levid  \lsep \hlcc{ \memory \lsep \datainst})\hist{: h}
        \to \\
        (c  \lsep \hlc{\writeBs{n};\task}) \hsep (\levid  \lsep   \hlcc{\memory  \setto{n}{\bot} \lsep  \datainst\msunion\fetchs{n}}) \hist{: h}
        }
$$
This rule handles the case when a core tries to write to a memory
block with address~$n$, which is either invalid or not found in the
first level cache.  The core will then be blocked while the memory
block is fetched from the lower level cache or from the main memory.
On the other hand, according to the annotation, no transition rule
corresponds to the execution path that is described by the negation of
the condition.  Note that this covers the case when the status
returned by \Abs{getStatus} (Line~\ref{lst:run.useWrite}) has changed;
i.e., the status of the memory block is no longer undefined or
invalid. As above, the \Abs{run} method terminates without having
successfully processed the \Abs{write(n)} task, which will be
evaluated again by the next asynchronous invocation of
the \Abs{run} method.

In the next section we show how to formally validate the annotations
in terms of a simulation relation.


\section{The Simulation Relation}
\label{sec: correctness}
We now establish the correctness of the ABS implementation of the
multicore memory system with respect to the multicore TSS specification.
First we observe that the structural correspondence described in
Section~\ref{sec:model} only relates the class diagram of the ABS
program (Figure~\ref{fig:classDiagram}) and the syntax of the runtime
configurations (Figure~\ref{fig:rts}) of the multicore TSS.  To relate
\emph{behavioral} information, we define 
the abstraction function $\alpha$ which maps every stable global ABS
configuration $G$ to a structurally equivalent configuration
$\alpha(G)$ which additionally provides a \emph{one-to-one mapping}
between the \emph{observable} processes of the instances of the ABS
class \Abs{Cache} and the \emph{dst} instructions of the corresponding
TSS cache representation \emph{Ca}, such that the actual address of
the associated \emph{dst} instruction equals the value of the
corresponding formal parameter of the ABS process.  The observable ABS
processes are those that stem from an asynchronous call of a method
that corresponds with a \emph{dst} instruction (like
\Abs{fetch}, etc.).

We have the following main theorem stating that the ABS multicore program is a correct
implementation of the multicore TSS as an instance of
Theorem~\ref{theorem:simulation-relation-methodology} (recall that
$\Rightarrow$ denotes the transition relation between stable ABS
configurations):

\begin{theorem}
  \label{theorem:simulation-relation}
  Let $G$ be a {reachable} stable global configuration of the ABS
  multicore model.  
If $G\Rightarrow G'$ then $\alpha(G)=\alpha(G')$
  or $\alpha(G)\rightarrow\alpha(G')$.
\end{theorem}

\noindent
\textbf{Proof of Theorem~\ref{theorem:simulation-relation}.}  
Because of the general confluence property of the ABS
semantics, it suffices to verify the annotations of the \Abs{run}
method and the methods of the \Abs{Cache} class that correspond to the \emph{dst} instructions
in terms of the simulation relation $\alpha$.   Here we detail the analysis
of the \Abs{Cache} class.

We first verify the annotations of the \Abs{fetch} method
(Figure~\ref{fig:abs-fetch}), identifying stable points
by their line numbers.
The \Abs{fetch}  method involves 
synchronous calls to the auxiliary
methods  \Abs{broadcast} (Figure~\ref{fig:abs-fetch}) and \Abs{swap} (Figure~\ref{fig:abs-swap}).
The method \Abs{broadcast} describes an instance of the global
synchronization pattern (Figure~\ref{fig:absglobalsync}).
The method \Abs{sendRd} of the \Abs{bus} asynchronously calls the
method \Abs{receiveRd}, see Figure~\ref{fig:abs_receiveRd}, of all
caches (except for the calling cache), using the barrier
synchronization (again, see Figure~\ref{fig:absglobalsync},
Section~\ref{sec:abs}).
The \Abs{swap} method is an instance of the test-and-set pattern, shown in
Figure \ref{fig:TAS}.

\begin{figure}[t]
\input{AnnotatedABS/fetch_abs}
\caption{\label{fig:abs-fetch} The annotated \Abs{fetch}  method.}
\end{figure}

\bigskip  
\begin{figure}[t]
\input{AnnotatedABS/swap}
\caption{\label{fig:abs-swap}The \Abs{swap}  method.}
\bigskip
\input{AnnotatedABS/receiveRd_abs}
\caption{\label{fig:abs_receiveRd}The annotated  \Abs{receiveRd} method.}
\end{figure}

Let $G\Rightarrow G'$ describe the execution of an invocation of the
\Abs{fetch} method from one stable point to a next one (or to termination), and let
$\levid \lsep \memory \lsep \datainst\msunion\fetchs{n}$ be the cache
in $\alpha(G)$ that corresponds to the cache in $G$ executing the
\Abs{fetch} method, where $n$ denotes the value of the formal
parameter \Abs{n} of the executing invocation of the \Abs{fetch}
method.  Further, let $\levid' \lsep \memory' \lsep \datainst'$ be the
cache in $\alpha(G)$ that corresponds to the next level cache in
$\alpha(G)$, if defined.  If such a next level cache exists, we have
that $\levno{\levid'} = \levno{\levid} + 1$ and
$\coreid{\levid} = \coreid{\levid'}$.  By $\overline{M}$ we denote in
the following the \emph{main} memory in $\alpha(G)$.  We have the
following case analysis of  $G\Rightarrow G'$ that describe the
execution of an invocation of the \Abs{fetch} method from one stable
point to a next one (or to termination).

\emph{Lines~\ref{lst:fetch.begin} $\Rightarrow$
\ref{lst:fetch.nl.nothing.in}.} In this case, $G\Rightarrow G'$
involves the execution of the \Abs{fetch} method (by a cache object)
starting from the beginning of the method (Line~\ref{lst:fetch.begin})
to the synchronous call of the method \Abs{remove\_inv}
(Line~\ref{lst:fetch.nl.nothing.in}) of the next level cache (note
that thus \Abs{nextCache} \textsf{!=} \Abs{Nothing} holds).  Since this invalidates the path condition of  the  annotation
associated with the beginning of the \Abs{fetch} method, by definition (see
Section~\ref{sec:Anno}), there is \emph{no} transition rule (of the
multicore TSS) which corresponds to $G\Rightarrow G'$.  This
execution only adds this call to the queue of the next level cache and
$\alpha$ abstracts from invocations of the method \Abs{remove\_inv}, so
it follows that $\alpha(G')=\alpha(G)$.

\emph{Lines~\ref{lst:fetch.nl.nothing.in} $\Rightarrow$
\ref{lst:fetch.nl.nothing.in.tfb}.} In this case, $G\Rightarrow G'$
consists of the path which leads from the execution of the called
\Abs{remove\_inv} method via the \Abs{then}-branch of the subsequent
\Abs{if}-statement to the termination of the \Abs{fetch} method
(Line~\ref{lst:fetch.nl.nothing.in.tfb}).  According to the annotation
\Abs{removed==true} : $\crule{\rn{LC-Miss}}$, this execution path
should correspond to the following application of the
$\crule{\rn{LC-Miss}}$ rule:
$$
 \ncondrule{LC-Miss}{
	\levno{\levid'} = \levno{\levid} + 1 \quad \coreid{\levid} = \coreid{\levid'} \quad
	 \statusfunc{\memory'}{n} \in \{\inv,\bot\} 
      }
      {
	(\levid \lsep \memory \lsep \hlc{\datainst\msunion\fetchs{n}})\hsep(\levid' \lsep \hlcc{\memory' \lsep \datainst'})
	\to \\
	(\levid \lsep \memory \lsep \hlc{\datainst\msunion\blocks{n}})\hsep(\levid'\lsep \hlcc{\memory' \setto{n}{\bot} \lsep \datainst'\msunion\fetchs{n}})
      }
$$
This rule handles the situation where a cache is trying to fetch a
memory block~$n$ from its next level cache, but the block is either
invalidated or does not exist in the cache.  The \Abs{fetch}-method in
this cache will then be suspended; \Abs{fetch} is propagated to the
next level cache and the memory block~$n$ will be removed from the
next level cache.
Since in this case the method
\Abs{remove\_inv} has returned the Boolean value ``True'' we can infer
statically from its code (Figure~\ref{fig:abs-aux},
Section~\ref{sec:model}) that the initial status of the given address
in the next level cache is \Abs{Nothing} or \Abs{Just(In)}.
Thus, the conditions for this application of the \rn{LC-Miss} rule are
enabled in $\alpha(G)$.
Moreover, we can statically infer in this case that the execution of
the \Abs{fetch} method is simulated by the updates
$\datainst\msunion\blocks{n}$ and $\datainst'\msunion\fetchs{n}$, and
the \Abs{remove\_inv} method is simulated by the update
$\memory' \setto{n}{\bot}$.  We conclude that
$\alpha(G)\rightarrow \alpha(G')$ by this application of the
$\crule{\rn{LC-Miss}}$ rule.

\emph{Lines~\ref{lst:fetch.nl.nothing.in} $\Rightarrow$
  \ref{lst:fetch.nl.sh.mo.swap}.} In this case, $G\Rightarrow G'$
consists of the path which leads from the execution of the
\Abs{remove\_inv} method via the \Abs{else}-branch of the subsequent
\Abs{if}-statement to the (synchronous) call of the \Abs{swap} method
(Line~\ref{lst:fetch.nl.sh.mo.swap}).  Since in this case the method
\Abs{remove\_inv} has returned the Boolean value ``False'', the
configuration $G'$ results from $G$ by initializing the local variable
\Abs{selected} and queuing the call of the \Abs{swap} method.
Abstracting from the definition of the ABS \Abs{select} function,
which picks a cache line for eviction to give space to a newly fetched
memory block, we simply assume that the first element of the pair
denoted by the ABS expression \Abs{select(cacheMemory,n)} equals the
address $\selectfunc{\memory}{n}$ and its second element equals the
status of this address.\footnote{In the actual ABS implementation, an
  extra parameter is used to capture the maximum size of the cache to
  check if there is free space to fetch a memory block $n$ from its
  next level.}  By definition of the simulation relation which
abstracts from local variables and invocations of auxiliary methods
like the \Abs{swap} method, it follows that $\alpha(G')=\alpha(G)$.

\emph{Lines~\ref{lst:fetch.nl.sh.mo.swap} $\Rightarrow$
  \ref{lst:fetch.nl.sh.mo.nn.put}.} In this case, $G\Rightarrow G'$
describes one of the two  paths which lead from the  execution of 
the \Abs{swap} method via the \Abs{then}-branch of the subsequent
\Abs{if}-statement (so \Abs{s} \textsf{!=} \Abs{Nothing}) to the
termination of the \Abs{fetch} method
(Line~\ref{lst:fetch.nl.sh.mo.nn.put}).  
We  infer statically from \Abs{s} \textsf{!=} \Abs{Nothing}
that $ \memory'(n) = \pr{k_j}{s'}$, where $s' = \sh \lor s' = \mo $.
Further, because $G$ is reachable we infer from the semantics of
synchronous calls that \Abs{selected==select(cacheMemory,n)}
holds in $G$ for the given cache object executing the \Abs{fetch} method.
Finally, we observe that
\Abs{fst(selected) == n} holds at Line~\ref{lst:fetch.sh.mo.nn} of the
\Abs{fetch} method if and only if \Abs{fst(n\_in) == n\_out} holds at
Line~\ref{lst:swap.sh.mo.nin} of the \Abs{swap} method. There are two
cases, depending on the value of \Abs{fst(selected)}.

First, let \Abs{fst(selected)} \textsf{!=} \Abs{n}, that is,  $ \selectfunc{\memory}{n}\not=n$.
It follows that  the enabling conditions of the
$\crule{\rn{LC-Hit}_1}$ rule
$$
	\ncondrule{LC-Hit$_1$}{
	\levno{\levid'} = \levno{\levid} + 1 \quad \coreid{\levid} = \coreid{\levid'} \\
        \selectfunc{\memory}{n} = m   \quad n\neq m\quad
        \memory(m) = \pr{k_i}{s} \quad \memory'(n) = \pr{k_j}{s'} \quad s'  = \sh  \lor s' = \mo 
      }
      {
	(\levid  \lsep \hlc{\memory} \lsep \datainst\msunion\fetchs{n})\hsep(\levid' \lsep \hlcc{\memory'} \lsep \datainst')
	\to \\
	(\levid  \lsep \hlc{\memory[m\mapsto \bot,n\mapsto
          \pr{k_j}{s'}]} \lsep \datainst) \hsep 
	(\levid' \lsep \hlcc{\memory' [n\mapsto \bot,m\mapsto \pr{k_i}{s}]} \lsep \datainst')
      }
$$
are satisfied in $\alpha(G)$.  Rule \rn{LC-Hit$_1$} addresses the case
where a cache finds~$n$, i.e., the block to be fetched, in its next
level with status shared or modified, and there is no free space in
the cache to place~$n$.  In order to fetch~$n$, the rule selects a
memory block~$m$ in the current cache to be swapped with~$n$ in the
next level cache.
According to the annotation \Abs{s} \textsf{!=} \Abs{Nothing}
\textsf{\&} \Abs{fst(selected)} \textsf{!=} \Abs{n} :
$\crule{\rn{LC-Hit}_1}$, this rule should correspond to the path
identified by its condition.  Since \Abs{fst(n\_in)} \textsf{!=}
\Abs{n\_out}, the execution of the \Abs{swap} method by the next level
cache in ABS is simulated by the update
$\memory' [n\mapsto \bot, m\mapsto \pr{k_i}{s}]$ and, on the other
hand, the execution of the two assignments (Lines
$\ref{lst:fetch.nl.sh.mo.nn.select.nn.remove}$ and
\ref{lst:fetch.nl.sh.mo.nn.put} of the \Abs{fetch} method) is simulated
by the update $\memory[m\mapsto \bot,n\mapsto \pr{k_j}{s'}]$.

Next, let \Abs{fst(selected)==n}, that is,
$ \selectfunc{\memory}{n}==n$.  It follows that the enabling
conditions of the $\crule{\rn{LC-Hit}_2}$ rule
$$
 \ncondrule{LC-Hit$_2$}{
	\levno{\levid'} = \levno{\levid} + 1 \quad \coreid{\levid} = \coreid{\levid'} \\ \selectfunc{\memory}{n} = n \quad
	\memory'(n) = \pr{k_j}{s'} \quad s'  = \sh  \lor s' = \mo 
      }
      {
	(\levid   \lsep \hlc{\memory} \lsep \datainst\msunion\fetchs{n})\hsep(\levid' \lsep \hlcc{\memory'} \lsep \datainst')
	\to \\
	(\levid  _i \lsep \hlc{\memory[n\mapsto \pr{k_j}{s'}]} \lsep \datainst)\hsep 
	(\levid' \lsep \hlcc{\memory' [n\mapsto \bot]} \lsep \datainst')
      }
$$
are satisfied. This rule addresses the case where a cache finds~$n$,
i.e., the memory block to be fetched, in its next level cache with
status shared or modified, and there is free space in the current
cache to place~$n$.  According to the annotation \Abs{s} \textsf{!=}
\Abs{Nothing} \textsf{\&} \Abs{fst(selected) == n :}
$\crule{\rn{LC-Hit}_2}$, this rule should correspond to the the path
identified by its condition.  Since \Abs{fst(n\_in) == n\_out}, in
this case the execution of the \Abs{swap} method by the next level
cache in ABS is simulated by the $\memory' [n\mapsto \bot]$ update
and, on the other hand, the execution of the assignment
(Line~\ref{lst:fetch.nl.sh.mo.nn.put} of the \Abs{fetch} method) is
simulated by the update $\memory[n\mapsto \pr{k_j}{s'}]$.

\emph{Lines~\ref{lst:fetch.nl.sh.mo.swap} $\Rightarrow$
  \ref{lst:fetch.nl.sh.mo.nn}.} In this case, $G\Rightarrow G'$
describes the execution path which leads from the return of the
\Abs{swap} method via the \Abs{else}-branch of the subsequent
\Abs{if}-statement (so \Abs{s == Nothing}) to the termination of the
\Abs{fetch} method.  We infer statically from the code of the
\Abs{swap} method and the condition \Abs{s == Nothing} that the status
of the address denoted by the formal parameter \Abs{n} of the
\Abs{fetch} method of the next level cache is undefined or invalid.
According to the annotation of the \Abs{swap} method, no rule of the
multicore TSS is applicable.  Since this invocation of the \Abs{fetch}
method terminates after an asynchronous self-call transmitting the
same address, we have that $\alpha(G)=\alpha(G')$.

\emph{Lines~\ref{lst:fetch.begin} $\Rightarrow$
  \ref{lst:fetch.ll.tfb}} In this case, $G\Rightarrow G'$ involves the
execution of the \Abs{fetch} method starting from the beginning of the
method which leads to
 termination after the execution of the
\Abs{broadcast} method  and the asynchronous self-call to
the \Abs{fetchBl} method.  The method \Abs{broadcast} implements an
instance of the global synchronization pattern (Figure~
\ref{fig:absglobalsync}).  The synchronous call to the \Abs{sendRd}
method of the bus generates asynchronous calls to the \Abs{receiveRd}
method (Figure~\ref{fig:abs_receiveRdx}) of all caches except the one
that initiated the global bus synchronization.  According to the
annotation $\crule{\rn{LLC-Miss/Synch}}$, this execution path
corresponds to the global synchronization rule
$$
 \ncondrule{Synch}{

        \many{\cache}\hist{: H}
          \toL{\red{\sendR{n}}} 
        \many{\cache'}\hist{: H'} 

        }{
        \langle \many{\core}  \hsep \many{\cache} \hsep \memory \rangle \hist{: H}
        \to
        \langle \many{\core}  \hsep \many{\cache'} \hsep \memory \rangle \hist{: H'}}
$$
where the premise is generated by successive applications of the rule
$$
\ncondrule{Synch-Dist}{
        \cache_1\not\in\many\cache\quad
        \many{\cache} 
        \toL{\red{\sendR{n}}} 
         \many{\cache'} 
        \quad
        \cache_1
        \toL{\red{\recR{n}}}
       \cache'_2
        }{
        \many{\cache}\cup\{ \cache_1\}
       \toL{\red{\sendR{n}}} 
         \many{\cache'}\cup \{ \cache_2\}
        } 
$$
This latter rule itself is  triggered by the rules
$$
          \ncondrule{LLC-Miss}{
	\last{\levid} = true 
	}
      {
	(\levid  \lsep \memory \lsep \hlc{\datainst\msunion\fetchs{n}})  \toL{\red{\sendR{n}}} 
        (\levid  \lsep \memory \lsep \hlc{\datainst\msunion\blocks{n}})
      }   
$$
and 
$$ 
  \begin{array}[t]{c}
      \ncondrule{Flush-One-Line}{

        \statusfunc{\memory}{n}  = \mo 
      }{
        \levid  \lsep \memory \lsep  \hlc{\datainst} 
        \toL{\red{\recR{n}}} 
        \levid  \lsep \memory \lsep  \hlc{\datainst\msunion\flushs{n}} 
      }
      
 \qquad
      
      \ncondrule{Ignore-Flush-One-Line}{
       \statusfunc{\memory}{n} \not = \mo 
     }{
       \levid  \lsep \memory \lsep  \datainst 
       
       \toL{\red{\recR{n}}} 
       \levid  \lsep \memory \lsep  \datainst 
     }
     
%

     \end{array}
$$
Together, these rules capture the broadcast mechanism for getting the
most recent shared copy of a memory block in the multicore memory
system.  Rule \rn{LLC-Miss} corresponds to the case when a last level
cache needs to fetch a memory block~$n$,
by broadcasting a read message, $\sendR{n}$,
to all other caches.  This is achieved through triggering the global
synchronisation rules~\rn{Synch} and \rn{Synch-Dist}.  While the
former identifies the core~$\core$
that broadcasts the read message, the latter propagates
the message~$\sendR{n}$
to the other caches.  Depending on the status of block~$n$,
the recipient cache will either flush the local modified copy to the
main memory (\rn{Flush-One-Line}) or ignore the message
(\rn{Ignore-Flush-One-Line}).

Clearly, the \Abs{receiveRd} method is simulated by the above rules
$\crule{\rn{Flush-One-Line}}$
and $\crule{\rn{Ignore-Flush-}}$ $\crule{\rn{One-Line}}$.
Since we may assume (as argued in Section \ref{sec:abs}) that the
execution of the \Abs{broadcast} method only consists of an
interleaving of the processes that are generated by it, it is easy to
statically verify that the execution of
\Abs{this.broadcast(this,n);this}!\Abs{fetchBl(n)} is simulated by the
above rules.

What remains is the correctness of the methods \Abs{fecthBl},
\Abs{fetchW}, and \Abs{flush} (see Figures~\ref{fig:abs-fetchbl},
\ref{fig:abs-fetchWl}, and \ref{fig:abs-flush}).  This can be
established in a similar manner as the above correctness proof for the
\Abs{fetch} method. Below we discuss these correctness proofs,
omitting details which are straightforward to check.

The condition of the annotation at the beginning of the \Abs{fetchBl}
method identifies the path which terminates after the call of the
\Abs{fetchW} method (Line~\ref{fetchBl.fetchW}). It is straightforward
to check that this path is simulated by the rule
$$
   \ncondrule{FetchBl$_3$}{
       
      \last{\levid} = true \quad \selectfunc{\memory}{  n} = n'  \quad n' \not = n \quad
      \statusfunc{\memory}{n'}  =  \mo \quad 
    }{
      (\levid  \lsep \memory \lsep \hlc{\datainst\msunion\blocks{n}}) 
      
      \to
      (\levid  \lsep \memory \lsep \hlc{\datainst\msunion\flushs{n'} \msunion\fetchwait{n}{n'}}) 
    }  
$$
which handles the case where a last level cache is blocked on~$n$
and the location where~$n$
is to be placed is occupied by a modified memory block.  To free the
location for~$n$,
the modified block needs to be flushed to the main memory by the
\key{flush} instruction, and the fetching will continue with the
\key{fetchW} instruction.
By definition, no rules correspond to execution paths starting from
the beginning of the \Abs{fetchBl} method and leading to the
synchronous calls of the \Abs{getStatus} method
(Lines~\ref{fetchBl.statusnext}, \ref{fetchBl.statusmain1}, and
\ref{fetchBl.statusmain2}).

The condition of the annotation \Abs{status} \textsf{!=} \Abs{Nothing}
: $\crule{\rn{LLC-Fetch-Unblock}}$
identifies the path which leads from the execution of the \Abs{getStatus}
method of the next level cache (Line~\ref{fetchBl.statusnext})  and
subsequent execution of the \Abs{else}-branch of
the subsequent \Abs{if}-statement to termination. It is straightforward to check that
this path is simulated by the rule
$$
   \ncondrule{LC-Fetch-Unblock}{
	\levno{\levid'} = \levno{\levid} + 1 \quad \coreid{\levid} = \coreid{\levid'} \quad n \in \dom{\memory'}
           }
      {
	(\levid \lsep \memory \lsep \hlc{\datainst\msunion\blocks{n}})\hsep(\levid' \lsep \memory' \lsep \datainst')
	\to  \\
	(\levid \lsep \memory \lsep \hlc{\datainst\msunion\fetchs{n}})\hsep 
	(\levid'\lsep \memory' \lsep \datainst')
      }   
$$
which corresponds to the case where~$n$,
i.e., the memory block that the cache is blocked on, is found in the
next level cache.  The rule unblocks the cache by trying to
fetch~$n$
again (replacing $\blocks{n}$
with $\fetchs{n}$
in $\datainst$).
By definition, no rule corresponds to the path which leads to
termination after execution of the \Abs{then}-branch.  In this case
the execution of the \Abs{fetchBl} method involves busy waiting for
the status of the given address in the next level cache to become
defined.

The annotation $\crule{\rn{FetchBl}_1}$ identifies the path which leads
from execution of  the \Abs{getStatus} method of the main memory and
the
update \Abs{cacheMemory = put(cacheMemory,n,status)}
(Line~\ref{fetchBl.statusmain1} to termination.  It is
straightforward to check that this path is simulated by the rule
$$
   \ncondrule{FetchBl$_1$}{
      \last{\levid} = true \quad  \selectfunc{\memory}{n} = n
      
       \quad
      s =  \statusfunc{\overline{\memory}}{n}  
    }{
     
      (\levid  \lsep \hlc{\memory\lsep \datainst\msunion\blocks{n}}) \hsep  \overline{\memory}   \to 
      ( \levid  \lsep \hlc{\memory'\setto{n}{ \pr{k}{s}} \lsep   \datainst
      }) \hsep  \overline{\memory} 
    } 
$$
which unblocks a last level cache which has space to fetch the memory
block~$n$
from main memory~$\many{M}$.
Similarly, the annotation $\crule{\rn{FetchBl}_2}$
identifies the path  which leads from execution of the  \Abs{getStatus} method
of the main memory (Line~\ref{fetchBl.statusmain2})  and the subsequent  updates \Abs{removeKey(cacheMemory,selected\_n)}
and \Abs{cacheMemory =} \Abs{put(cacheMemory,n,status)} to termination. It is
straightforward to check that this path is simulated by the rule
$$
 \ncondrule{FetchBl$_2$}{
      \last{\levid} = true \quad \selectfunc{\memory}{n} = n'
       \quad n' \not = n  \quad
      \statusfunc{\memory}{n'}  \not = \mo
      \quad  
       s = \statusfunc{\overline{\memory}}{n}   
    }{
     
      (\levid  \lsep \hlc{\memory'\lsep \datainst\msunion\blocks{n}})  \hsep \overline{\memory}
      \to
      (\levid  \lsep \hlc{\memory'[n'\mapsto \bot, n \mapsto  \pr{k}{s}] \lsep \datainst
      }) \hsep  \overline{\memory} 
    } 
$$
which corresponds to a last level cache fetching the memory
block~$n$
from main memory, but the location where~$n$
is to be placed is occupied by a non-modified block. The rule then
removes the occupying block and places~$n$ into the cache.

\begin{figure}[t]
\input{AnnotatedABS/fetchBl_abs}
\caption{\label{fig:abs-fetchbl}The annotated \Abs{fetchBl} method.}
\end{figure}

Concerning the \Abs{fetchW} method (Figure~\ref{fig:abs-fetchWl}), it
is straightforward to check that the path which leads from
the \Abs{await} statement to its termination is simulated by the rule.

$$
  \ncondrule{FetchW}{
       
      \last{\levid} = true   \quad
      \statusfunc{\memory}{n'}  \neq  \mo \quad 
    }{
      (\levid  \lsep \memory \lsep \hlc{\datainst\msunion\fetchwait{n}{n'}}) 
      
      \to
      (\levid  \lsep \memory \lsep \hlc{\datainst\msunion\blocks{n}}) 
    }  
$$
which handles the case where a last level cache, which fails to
replace~$n'$ with~$n$
because~$n'$ was a modified block, can try to fetch~$n$ again.

\begin{figure}[h!]
\input{AnnotatedABS/fetchblw_abs}
\caption{\label{fig:abs-fetchWl}The annotated \Abs{fetchW}  method.}
\end{figure}

\begin{figure}[h!]
\input{AnnotatedABS/flush_abs}
\caption{\label{fig:abs-flush}The annotated \Abs{flush} method.}
\end{figure}

The path identified by the condition of the  annotation
\Abs{lookup(cacheMemory,n)} \textsf{!=} \Abs{Mo} : $\crule{\rn{Flush}_2}$
of the \Abs{flush} method (Figure~\ref{fig:abs-flush})
 is clearly simulated by the  rule 
$$
   \ncondrule{Flush$_2$}{
       \statusfunc{\memory}{n}  \neq \mo
    }{
     (\levid  \lsep \memory \lsep  \hlc{\datainst\msunion\flushs{n}}) 
      \to 
      (\levid  \lsep \memory \lsep   \hlc{\datainst})
    }  
$$
which just ignores a \key{flush} instruction if the block is not
modified.  By definition of this annotation, no rule corresponds with
the path leading to the synchronous call of the method \Abs{setStatus}
of \Abs{mainMemory}.  According to the annotation
$\crule{\rn{Flush}_1}$, the path which leads from execution of 
the \Abs{setStatus} method of the main memory  to
termination, corresponds to the rule 
$$
 \ncondrule{Flush$_1$}{
        \statusfunc{\memory}{n}  = \mo
    }{
     (\levid  \lsep \hlc{\memory \lsep  \datainst\msunion\flushs{n}}) \hsep    \hlcc{\overline{\memory}} 
      \to 
      (\levid  \lsep \hlc{\memory\setto{n}{ \pr{k'}{\sh}} \lsep  \datainst}) \hsep  \hlcc{\overline{\memory}\setto{n}{ \pr{k'}{\sh}}} 
    }  
$$
which flushes a modified memory block~$n$ by updating its status to
shared in both the cache and the main memory.  Clearly, by the
simulation relation, the condition of this rule is satisfied in
$\alpha(G)$.  Further, the update \Abs{cacheMemory =
  put(cacheMemory,n,Sh)} is simulated by
$\memory\setto{n}{ \pr{k'}{\sh}}$ and the update
\Abs{mainMemory.setStatus(n,Sh)} is simulated by
$\overline{\memory}\setto{n}{ \pr{k'}{\sh}}$.

\subsection{Bisimulation}
We briefly discuss how to extend Theorem~\ref{theorem:simulation-relation}
to a bisimulation between the transitive, reflexive closure of the transition
relation $\Rightarrow$ of the ABS multicore program and  that of the transition relation $\rightarrow$ of the multicore TSS.
Such a bisimulation relation then allows to prove both \emph{safety} and
\emph{liveness} properties of the ABS multicore program in terms of the multicore TSS.
The following theorem states that the multicore TSS is simulated by the ABS program.

\begin{theorem}
  \label{theorem:bisimulation}
  Let $G$ be a {reachable} stable global configuration of the ABS
  multicore program.  
If $\alpha(G)\rightarrow \mbox{\it cf}$ then there exists a stable  
 global configuration $G'$ such that $\alpha(G')=\mbox{\it cf}$ and 
$G\Rightarrow^* G'$.
\end{theorem}

\begin{proof}
{\rm 
We sketch a proof of this theorem which is  based on the correctness of the annotations as established in the proof of Theorem ~\ref{theorem:simulation-relation}.
The global structure of the proof of Theorem~\ref{theorem:bisimulation} however involves an analysis of the individual TSS rules.
All these rules are triggered by a \emph{dst} instruction.
For each such instruction we check statically for each stable point of the corresponding ABS method whether there exists a path to another stable
point (\emph{not necessarily the next one}) execution of which corresponds
to the TSS rule application.

As an example of this scheme we give an analysis of an application of the  rule 
$$
 \ncondrule{LC-Miss}{
	\levno{\levid'} = \levno{\levid} + 1 \quad \coreid{\levid} = \coreid{\levid'} \quad
	 \statusfunc{\memory'}{n} \in \{\inv,\bot\} 
      }
      {
	(\levid \lsep \memory \lsep \hlc{\datainst\msunion\fetchs{n}})\hsep(\levid' \lsep \hlcc{\memory' \lsep \datainst'})
	\to \\
	(\levid \lsep \memory \lsep \hlc{\datainst\msunion\blocks{n}})\hsep(\levid'\lsep \hlcc{\memory' \setto{n}{\bot} \lsep \datainst'\msunion\fetchs{n}})
      }
$$
in $\alpha(G)$, triggered by the \emph{dst}
instruction  $\fetchs{n}$.
By definition of $\alpha$ there exists 
a process instance
 (either executing or suspended) of the \Abs{fetch} method with the address  $n$ as the value of the its formal
parameter. 
Further, the status of the address $n$ of the next level cache, denoted by the \Abs{nextCache} field  of the  cache object to which
this process belongs, is undefined or invalid.
We have the following straightforward analysis of the stable points of this process.

In case of the initial stable point 
and the stable point associated with the call of the 
\Abs{remove\_inv} method (Line~\ref{lst:fetch.nl.nothing.in} of the \Abs{fetch} method, Figure~\ref{fig:abs-fetch}),  by definition of $\alpha$ there exists a computation $G\Rightarrow^* G'$
which involves in both cases  the execution of the path from 
Line~\ref{lst:fetch.nl.nothing.in}
to Line \ref{lst:fetch.nl.nothing.in.tfb}
As argued in the proof of Theorem ~\ref{theorem:simulation-relation}, this path corresponds to an application of the 
 $\crule{\rn{LC-Miss}}$ rule.
Note that execution of the path from Line~\ref{lst:fetch.begin}
to  Line~\ref{lst:fetch.nl.nothing.in} corresponds to a 
silent transition in the multicore TSS,

In case of the stable point associated with the call of the  \Abs{swap} method
(Line ~\ref{lst:fetch.nl.sh.mo.swap}),  by definition of $\alpha$ there exists a computation $G\Rightarrow^* G'$ 
which involves execution of the  path which leads from the return of the
\Abs{swap} method via the \Abs{else}-branch of the subsequent
\Abs{if}-statement (note that  \Abs{s == Nothing}) to  termination of the
\Abs{fetch} method, \emph{followed} by execution of the  path  from the initial stable point of the \emph{newly} generated process instance of the \Abs{fetch} method
to Line \ref{lst:fetch.nl.nothing.in.tfb}. As argued above, this latter  path corresponds to an application of the 
 $\crule{\rn{LC-Miss}}$ rule. Execution of  the first path corresponds to a 
silent transition in the multicore TSS (as argued in the proof of Theorem ~\ref{theorem:simulation-relation}).

}
\end{proof}


\section{Related Work}
\label{sec:related}

There is in general a significant gap between a transition
specification and its implementation in a (high-level) parallel
programming language~\cite{SchlatteJMTY18}.  Transition system
specifications \cite{plotkin04jlap} succinctly formalize operational
models and are well-suited for proofs, but direct implementations of
such specifications quickly lead to very inefficient
implementations. Executable semantic frameworks such as
Redex~\cite{felleisen09redex}, rewriting
logic~\cite{meseguer13ic,meseguer07tcs}, and
$\mathbb{K}$~\cite{rosu17marktoberdorf} reduce this gap, and have been
used to develop executable formal models of complex languages like
C~\cite{ellison12popl} and Java~\cite{bogdanas15popl}.  The
relationship between transition system specifications and rewriting
logic semantics has been studied \cite{serbanuta09ic} without
proposing a general solution for synchronization by label
matching. Bijo et al.\ implemented their multicore memory model
\cite{bijo16wrla} in the rewriting logic system Maude \cite{maudebook}
using an orchestrator for label matching, but do not provide a
correctness proof wrt.\ the transition system specification. Different
semantic styles can be modeled and related inside one framework; for
example, the correctness of distributed implementations of KLAIM
systems in terms of simulation relations have been studied in
rewriting logic \cite{eckhardt15scp}.  Compared to these works on
semantics, we developed a general methodology for proving the
correctness of parallel implementations of transition system
specifications in the active object language ABS.  Our methodology
features a new integration of these two formalisms which consists of a
formal scheme for annotating ABS programs with transition rules.
These annotations provide a high-level specification of the proof
obligations for establishing the simulation relation between a
transition system specification and its ABS implementation.

Correctness-preserving compilation and refinement is related to
correctness proofs for implementations, and ensures that the low-level
representation of a program preserves the properties of the high-level
model. Examples of this line of work include the B-method
\cite{EventB}, which is based on refinement between abstract state
machines, type-preserving translations into typed assembly languages
\cite{morrisett99toplas}, and formally verified compilers
\cite{leroy09cacm,leroy09jar}, which proves the semantic preservation
of a compiler from C to assembler code, but leaves shared-variable
concurrency for future work.  In contrast to these works
 our work
 specifically
targets the correctness of parallel systems.

Simulation tools for cache coherence protocols can evaluate
performance and efficiency on different architectures (e.g.,
gems~\cite{martin05can} and gem5~\cite{binkert11can}).  These tools
perform evaluations of, e.g., the cache hit/miss ratio and response
time, by running benchmark programs written as low-level read and
write instructions to memory.  Advanced simulators such as
Graphite~\cite{miller*:graphite} and
Sniper~\cite{carlson.heirman.eeckhout:sniper} run programs on
distributed clusters to simulate executions on multicore architectures
with thousands of cores.  Unlike our work, these simulators are not
based on a formal semantics and correctness proofs.  Our work
complements these simulators by supporting the executable exploration
of design choices from a programmer perspective rather from hardware
design.  Compared to worst-case response time analysis for concurrent
programs on multicore architectures~\cite{li*:timeanalysis.multicore},
our focus is on the underlying data movement rather than the response
time.


%
\section{Conclusion}
\label{sec:conclusion}
We have introduced in this paper a methodology for proving the
correctness of parallel implementations of high-level transition
system specifications in the active object language ABS.  The proof
method consists of establishing a simulation relation between the
transition system describing the semantics of the ABS program and the
transition system described by the specification.  The proof method
exploits a general global confluence property of the ABS semantics
which allows to abstract from the interleaving of parallel processes
and focus on the static analysis of sequential code in the simulation
proof.  A promising further formalization and tool-supported
automation of our methodology is the \emph{symbolic execution} of
sequential ABS code in establishing the simulation relation between
the ABS program and its specification.

A concern that often arises in parallel execution is fairness: the
degree of variability when distributing the computing resources among
different parallel components --- here, the simulated cores. 
Fairness of parallel execution can affect the simulation's accuracy in
approximating the intended (or idealized) manycore hardware. To ensure
fairness of the simulation, we make use of \emph{deployment
components}~\cite{johnsen15jlamp} in ABS.

A \emph{Deployment Component} (DC) is an ABS execution location that
is created with a number of virtual resources (e.g., execution speed,
memory use, network bandwidth), which are shared among its deployed
objects. Any annotated statement \Abs{[Cost: x]}$ \ S$ decrements
by~\Abs{x} the resources of its DC and then completes, or it will
stall its computation if there are currently not enough resources
remaining; the statement~$S$ may continue on the next passage of the
global symbolic time where all the resources of the DCs have been
renewed, and will eventually complete when its \Abs{Cost} has reached
zero.

We make use of this resource modeling of ABS to assign equal (fair)
resources of virtual execution speed to the simulated cores of the
system. Each \Abs{Core} object is deployed onto a separate DC with
fixed \Abs{Speed(1)} resources. The processing of each instruction has
the same cost \Abs{[Cost: 1]} --- a generalization, since common
processor architectures execute different instructions in different
speeds (cycles per instruction); e.g., \verb+JUMP+ is faster than
\verb+LOAD+.  The result is that all \Abs{Core}s can execute maximum
one instruction in every time interval of the global symbolic clock,
and thus no \Abs{Core} can get too far ahead with processing its own
instructions --- a problem that manifests upon the parallel simulation
of~$N$ number of cores using a physical machine of~$M$ cores,
where~$N$ is vastly greater than~$M$.

We plan further development of this extension of the ABS multicore
model with deployment components for simulating the execution of
(object-oriented) programs on multicore architectures.  A first such
development concerns an extension of the abstract memory model with
data. In particular, having the addresses of the memory locations
themselves as data allows to model and simulate different data layouts
of the dynamically generated object structures.


\bibliographystyle{plain}

\appendix

\section{Multicore TSS}\label{sec:TSS}
The multicore TSS is structured in terms of separate TSS's for the
cores, caches, an global synchronization.  In general, we assume that
the unlabelled transitions which describe the behavior of the
individual cores and caches are applied in the context of a
configuration \emph{cf}, \red{and we omit the straightforward context rule
here}.  On the other hand, for the labelled transitions we introduce
explicit synchronization rules for lifting them to a particular
context.

\subsection*{Transition Rules for Cores}
Figure~
\ref{fig:local.sos.1}  shows the transition rules for 
the basic core  instructions $\reads{r}$, $\readBs{r}$, $\writes{r}$, and $\writeBs{r}$.

\begin{figure}[h]
      \centering
      { 
        \footnotesize
        \begin{displaymath}    
          \begin{array}[t]{c}
            \ncondrule{PrRd$_1$}{
            \quad   \first{\levid} = true   \quad \coreid{\levid} = c 
            \quad
            \statusfunc{\memory}{n} \in \{\sh,\mo\} 
            }{
            (c  \lsep \hlc{\reads{n};\task}) \hsep (\levid  \lsep \memory \lsep \datainst) \hist{: h}
            \to
            (c  \lsep \hlc{\task})  \hsep (\levid \lsep \memory \lsep \datainst) \hist{: h; R(c, n)} }
            \\[30pt]
            
            \ncondrule{PrRd$_2$}{
            \quad  \first{\levid} = true   \quad \coreid{\levid} = c 
            \quad
            \statusfunc{\memory}{n} \in \{\inv,\bot\} 
            }{
            (c  \lsep \hlc{\reads{n};\task}) \hsep   (\levid \lsep \hlcc{\memory \lsep \datainst})\hist{: h}
            \to \\
            (c  \lsep \hlc{\readBs{n};\task}) \hsep (\levid \lsep \hlcc{\memory  \setto{n}{\bot} \lsep  \datainst\msunion\fetchs{n}})  \hist{: h}
            }
      
         \\[30pt]
        
  \ncondrule{PrRd$_3$}{
          \first{\levid} = true \quad \coreid{\levid} = c 
          \quad
         n \in \dom{\memory}
        }{
         (c  \lsep \hlc{\readBs{n};\task}) \hsep  (\levid   \lsep \memory \lsep \datainst)\hist{: h}
          \to 
          (c  \lsep \hlc{\reads{n};\task}) \hsep  (\levid  \lsep \memory \lsep \datainst)  \hist{: h; R(c, n)}
        }
    \\[30pt]
       
   \ncondrule{PrWr$_1$}{
          \first{\levid} = true \quad \coreid{\levid} = c  
          \quad
          \statusfunc{\memory}{n}  = \mo
        }{
           (c  \lsep \hlc{\writes{n};\task}) \hsep  (\levid   \lsep \memory \lsep \datainst) \hist{: h}
          \to 
           (c  \lsep \hlc{\task}) \hsep  (\levid  \lsep \memory \lsep \datainst) \hist{: h; W(c, n)}
        }
        
      \\[30pt]
        
   \ncondrule{PrWr$_2$}{
         \first{\levid} = true   \quad \coreid{\levid} = c 
        \quad
         \statusfunc{\memory}{n} = \pr{k}{\sh}
        }{
        (c  \lsep \hlc{\writes{n};\task}) \hsep  (\levid   \lsep \hlcc{\memory} \lsep \datainst)\hist{: h}
        \toL{\red{\sendRX{n}}} 
         (c  \lsep \hlc{\task}) \hsep (\levid  \lsep \hlcc{\memory  \setto{n}{\pr{k}{\mo}}} \lsep \datainst) \hist{: h; W(c, n)}
        }        
    \\[30pt]
        
   \ncondrule{PrWr$_3$}{
         \first{\levid} = true   \quad \coreid{\levid} = c 
        \quad
        \statusfunc{\memory}{n} \in \{\inv,\bot\}         }{
        (c  \lsep \hlc{\writes{n};\task}) \hsep (\levid  \lsep \hlcc{ \memory \lsep \datainst})\hist{: h}
        \to \\
        (c  \lsep \hlc{\writeBs{n};\task}) \hsep (\levid  \lsep   \hlcc{\memory  \setto{n}{\bot} \lsep  \datainst\msunion\fetchs{n}}) \hist{: h}
        }

    \\[30pt]

          \ncondrule{PrWr$_4$}{
          \first{\levid} = true \quad \coreid{\levid} = c  
          \quad
         n \in \dom{\memory}
        }{
         (c  \lsep \hlc{\writeBs{n};\task})\hsep   (\levid   \lsep \memory \lsep \datainst) \hist{: h}
          \to
           (c  \lsep \hlc{\writes{n};\task}) \hsep (\levid  \lsep \memory \lsep \datainst) \hist{: h}
        }

    \end{array}
  \end{displaymath}
  \vspace{-10pt}
 \caption{ \label{fig:local.sos.1}  Transition rules for $\reads{r}$, $\readBs{r}$, $\writes{r}$, and $\writeBs{r}$.}
    }
\end{figure}

\subsection*{Transition Rules for Caches}
These rules are further structured in terms of separate TSS's for the individual \emph{rst} instructions (Figures~  \ref{fig:local.sos.2}, \ref{fig:fetchBl}, \ref{fig:fetchW}, and \ref{fig:flush}).

\begin{figure}[h]
  \centering
  { 
         \footnotesize
    \begin{displaymath}    
      \begin{array}[t]{c}

	\ncondrule{LC-Hit$_1$}{
	\levno{\levid'} = \levno{\levid} + 1 \quad \coreid{\levid} = \coreid{\levid'} \\
        \selectfunc{\memory}{n} = m   \quad n\neq m\quad
        \memory(m) = \pr{k_i}{s} \quad \memory'(n) = \pr{k_j}{s'} \quad s'  = \sh  \lor s' = \mo 
      }
      {
	(\levid  \lsep \hlc{\memory} \lsep \datainst\msunion\fetchs{n})\hsep(\levid' \lsep \hlcc{\memory'} \lsep \datainst')
	\to \\
	(\levid  \lsep \hlc{\memory[m\mapsto \bot,n\mapsto
          \pr{k_j}{s'}]} \lsep \datainst) \hsep 
	(\levid' \lsep \hlcc{\memory' [n\mapsto \bot,m\mapsto \pr{k_i}{s}]} \lsep \datainst')
      }

      \\[40pt]
      
 \ncondrule{LC-Hit$_2$}{
	\levno{\levid'} = \levno{\levid} + 1 \quad \coreid{\levid} = \coreid{\levid'} \\ \selectfunc{\memory}{n} = n \quad
	\memory'(n) = \pr{k_j}{s'} \quad s'  = \sh  \lor s' = \mo 
      }
      {
	(\levid   \lsep \hlc{\memory} \lsep \datainst\msunion\fetchs{n})\hsep(\levid' \lsep \hlcc{\memory'} \lsep \datainst')
	\to \\
	(\levid  _i \lsep \hlc{\memory[n\mapsto \pr{k_j}{s'}]} \lsep \datainst)\hsep 
	(\levid' \lsep \hlcc{\memory' [n\mapsto \bot]} \lsep \datainst')
      }

              \\[40pt]
              
 \ncondrule{LC-Miss}{
	\levno{\levid'} = \levno{\levid} + 1 \quad \coreid{\levid} = \coreid{\levid'} \quad
	 \statusfunc{\memory'}{n} \in \{\inv,\bot\} 
      }
      {
	(\levid \lsep \memory \lsep \hlc{\datainst\msunion\fetchs{n}})\hsep(\levid' \lsep \hlcc{\memory' \lsep \datainst'})
	\to \\
	(\levid \lsep \memory \lsep \hlc{\datainst\msunion\blocks{n}})\hsep(\levid'\lsep \hlcc{\memory' \setto{n}{\bot} \lsep \datainst'\msunion\fetchs{n}})
      }
 
         \\[40pt]
         
          \ncondrule{LLC-Miss}{
	\last{\levid} = true 
	}
      {
	(\levid  \lsep \memory \lsep \hlc{\datainst\msunion\fetchs{n}})  \toL{\red{\sendR{n}}} 
        (\levid  \lsep \memory \lsep \hlc{\datainst\msunion\blocks{n}})
      }   
      
    \\[40pt]

   \end{array}
  \end{displaymath}
  \vspace{-10pt}
 \caption{Transition rules for $\fetchs{n}$.}
  \label{fig:local.sos.2}
}
\end{figure}

\begin{figure}[h]
  \centering
  { 
     \footnotesize
    \begin{displaymath}    
      \begin{array}[t]{c}

   \ncondrule{FetchBl$_1$}{
      \last{\levid} = true \quad  \selectfunc{\memory}{n} = n
       \quad
      s =  \statusfunc{\overline{\memory}}{n}  
    }{
     
      (\levid  \lsep \hlc{\memory\lsep \datainst\msunion\blocks{n}}) \hsep  \overline{\memory}   \to 
      ( \levid  \lsep \hlc{\memory'\setto{n}{ \pr{k}{s}} \lsep   \datainst
      }) \hsep  \overline{\memory} 
    } 
    
         \\[30pt]
         
 \ncondrule{FetchBl$_2$}{
      \last{\levid} = true \quad \selectfunc{\memory}{n} = n'
       \quad n' \not = n  \quad
      \statusfunc{\memory}{n'}  \not = \mo
      \quad  
       s = \statusfunc{\overline{\memory}}{n}   
    }{
     
      (\levid  \lsep \hlc{\memory'\lsep \datainst\msunion\blocks{n}})  \hsep \overline{\memory}
      \to
      (\levid  \lsep \hlc{\memory'[n'\mapsto \bot, n \mapsto  \pr{k}{s}] \lsep \datainst
      }) \hsep  \overline{\memory} 
    } 

  \\[30pt]
    
   \ncondrule{FetchBl$_3$}{
       
      \last{\levid} = true \quad \selectfunc{\memory}{  n} = n'  \quad n' \not = n \quad
      \statusfunc{\memory}{n'}  =  \mo \quad 
    }{
      (\levid  \lsep \memory \lsep \hlc{\datainst\msunion\blocks{n}}) 
      
      \to
      (\levid  \lsep \memory \lsep \hlc{\datainst\msunion\flushs{n'} \msunion\fetchwait{n}{n'}}) 
    }  
  
  \\[30pt]

   \ncondrule{LC-Fetch-Unblock}{
	\levno{\levid'} = \levno{\levid} + 1 \quad \coreid{\levid} = \coreid{\levid'} \quad n \in \dom{\memory'}
           }
      {
	(\levid \lsep \memory \lsep \hlc{\datainst\msunion\blocks{n}})\hsep(\levid' \lsep \memory' \lsep \datainst')
	\to  \\
	(\levid \lsep \memory \lsep \hlc{\datainst\msunion\fetchs{n}})\hsep 
	(\levid'\lsep \memory' \lsep \datainst')
      }

 \end{array}
\end{displaymath}
}
\caption{Transition rules for $\blocks{n}$.  }
\label{fig:fetchBl}
\end{figure}

\begin{figure}[h]
  \centering
  { 
     \footnotesize
    \begin{displaymath}    
      \begin{array}[t]{c}
  \ncondrule{FetchW}{
       
      \last{\levid} = true   \quad
      \statusfunc{\memory}{n'}  \neq  \mo \quad 
    }{
      (\levid  \lsep \memory \lsep \hlc{\datainst\msunion\fetchwait{n}{n'}}) 
      
      \to
      (\levid  \lsep \memory \lsep \hlc{\datainst\msunion\blocks{n}}) 
    }  

 \\[30pt]
\end{array}
\end{displaymath}
}
\caption{Transition rule for $\fetchwait{n}{n'}$.  
  }
\label{fig:fetchW}
\end{figure}

\begin{figure}[t!]
  \centering
  { 
     \footnotesize
    \begin{displaymath}    
      \begin{array}[t]{c}
 \ncondrule{Flush$_1$}{
        \statusfunc{\memory}{n}  = \mo
    }{
     (\levid  \lsep \hlc{\memory \lsep  \datainst\msunion\flushs{n}}) \hsep    \hlcc{\overline{\memory}} 
      \to 
      (\levid  \lsep \hlc{\memory\setto{n}{ \pr{k'}{\sh}} \lsep  \datainst}) \hsep  \hlcc{\overline{\memory}\setto{n}{ \pr{k'}{\sh}}} 
    }  
    
   \\[30pt]
    
   \ncondrule{Flush$_2$}{
       \statusfunc{\memory}{n}  \neq \mo
    }{
     (\levid  \lsep \memory \lsep  \hlc{\datainst\msunion\flushs{n}}) 
      \to 
      (\levid  \lsep \memory \lsep   \hlc{\datainst})
    }  
\end{array}
\end{displaymath}
}
\caption{Transition rules for $\flushs{n}$.  
  }
\label{fig:flush}
\end{figure}

\subsection*{Transition Rules for Global Synchronization}
These rules are further structured in terms of a TSS 
for labelled transitions (Figure~ \ref{fig:local.sos.3}) and a TSS of rules
for matching these labelled transitions (Figure~ \ref{fig:synch}).

\begin{figure}[h]
  \centering
  { 
    \footnotesize
    \begin{displaymath}    
      \begin{array}[t]{c}

        \ncondrule{Invalidate-One-Line}{
          \statusfunc{\memory}{n}= \sh
        }{
          \levid  \lsep \hlc{\memory} \lsep  \datainst 
          
        \toL{\red{\recRX{n}}}
        \levid  \lsep \hlc{\memory\setto{n}{\inv}} \lsep   \datainst   
      }
        \qquad
      
      \ncondrule{Ignore-Invalidate-One-Line}{ 
       \statusfunc{\memory}{n} \in \{\inv,\bot\} 
      }{
        \levid  \lsep \memory \lsep  \datainst   
        
        \toL{\red{\recRX{n}}} 
        \levid  \lsep \memory \lsep  \datainst 
     }
     
   \\[30pt]
     
     \ncondrule{Flush-One-Line}{
        \statusfunc{\memory}{n}  = \mo    
      }{
        \levid  \lsep \memory \lsep  \hlc{\datainst} 
        \toL{\red{\recR{n}}} 
        \levid  \lsep \memory \lsep  \hlc{\datainst\msunion\flushs{n}} 
      }
      
 \qquad
      
      \ncondrule{Ignore-Flush-One-Line}{
       \statusfunc{\memory}{n} \not = \mo 
     }{
       \levid  \lsep \memory \lsep  \datainst 
       
       \toL{\red{\recR{n}}} 
       \levid  \lsep \memory \lsep  \datainst 
     }

       \end{array}
     \end{displaymath}
     \vspace{-10pt}
     \caption{Labelled input transitions.} 
     \label{fig:local.sos.3}
   }
\end{figure}  

\begin{figure}[h]
  \centering
  { 
     \footnotesize
    \begin{displaymath}    
      \begin{array}[t]{c}
\ncondrule{Synch-Dist}{
        \cache_1\not\in\many\cache\quad
        \many{\cache} 
        \toL{\red{\sendR{n}}} 
         \many{\cache'} 
        \quad
        \cache_1
        \toL{\red{\recR{n}}}
       \cache'_2
        }{
        \many{\cache}\cup\{ \cache_1\}
       \toL{\red{\sendR{n}}} 
         \many{\cache'}\cup \{ \cache_2\}
        } 
      

        \quad

   \ncondrule{Synch}{

        \many{\cache}\hist{: H}
          \toL{\red{\sendR{n}}} 
        \many{\cache'}\hist{: H'} 

        }{
        \langle \many{\core}  \hsep \many{\cache} \hsep \memory \rangle \hist{: H}
        \to
        \langle \many{\core}  \hsep \many{\cache'} \hsep \memory \rangle \hist{: H'}}

\\[30pt]
 
        \ncondrule{Synch-DistX}{
        \cache_1\not\in \many{\cache} \quad
        \core,\many{\cache} 
         \toL{\red{\sendRX{n}}} 
         \core', \many{\cache'} 
        \quad
        \cache_1
         \toL{\red{\recRX{n}}}
       \cache_2
        }{
        \core, \many{\cache}\cup \{\cache_1\}
        \toL{\red{\sendRX{n}}} 
         \core', \many{\cache'}\cup  \{\cache_2\}
        } 
      
 \qquad

   \ncondrule{SynchX}{
         
        \core\not\in \many{\core_1}\quad
        \core, \many{\cache}\hist{: H}
         \toL{\red{\sendRX{n}}} 
        \core', \many{\cache'}\hist{: H'} 

        }{
       \langle  \many{\core_1} \cup\{\core\} \hsep \many{\cache} \hsep \memory\hist{: H} \rangle \\
        \to
      \langle   \many{\core_1}\cup\{\core'\}  \hsep \many{\cache'} \hsep  \memory\setto{n}{\pr{\_}{\inv}}  \hist{: H'}\rangle }

 \end{array}
\end{displaymath}
}
\caption{Transition rules for global synchronization/broadcast.  
  }
\label{fig:synch}
\end{figure}


\newpage
\section{Multicore ABS}\label{sec:multicore.abs}
In this section, we collect all the methods of the ABS models that we have discussed
in the paper to provide a full view of the ABS implementation of the
multicore memory system.
\begin{figure}[t]
  \input{AnnotatedABS/runcore_abs}
  \caption{\label{fig:appendix.ABScore} The annotated \Abs{run} method.}
\end{figure}

\begin{figure}[h!]
\input{AnnotatedABS/ABSAux}
\caption{\label{fig:appendix.abs-aux}Methods \Abs{getStatus}, \Abs{remove_inv},   and \Abs{broadcastX} of class \Abs{Cache}.}
\bigskip
\input{AnnotatedABS/receiveRdx_abs}
\caption{\label{fig:appendix.abs_receiveRdx}  The annotated \Abs{receiveRdX}  method.}
\end{figure}

\begin{figure}[t]
\input{AnnotatedABS/fetch_abs}
\caption{\label{fig:appendix.abs-fetch} The annotated \Abs{fetch}  method.}
\end{figure}

\begin{figure}[t]
\input{AnnotatedABS/swap}
\caption{\label{fig:appendix.abs-swap}The \Abs{swap}  method.}
\end{figure}

\bigskip

\begin{figure}[t]
\input{AnnotatedABS/receiveRd_abs}
\caption{\label{fig:appendix.abs_receiveRd}The annotated  \Abs{receiveRd} method.}
\end{figure}

\begin{figure}[t]
\input{AnnotatedABS/fetchBl_abs}
\caption{\label{fig:appendix.abs-fetchbl}The annotated \Abs{fetchBl} method.}
\end{figure}

\begin{figure}[h!]
\input{AnnotatedABS/fetchblw_abs}
\caption{\label{fig:appendix.abs-fetchWl}The annotated \Abs{fetchW}  method.}
\end{figure}

\begin{figure}[h!]
\input{AnnotatedABS/flush_abs}
\caption{\label{fig:appendix.abs-flush}The annotated \Abs{flush} method.}
\end{figure}


\end{document}